\documentclass[english,aps,superscriptaddress]{revtex4}
\usepackage[T1]{fontenc}
\usepackage[latin9]{inputenc}
\usepackage{mathrsfs}
\usepackage{amsbsy}
\usepackage{amssymb}
\usepackage{esint}

\makeatletter
\@ifundefined{textcolor}{}
{%
 \definecolor{BLACK}{gray}{0}
 \definecolor{WHITE}{gray}{1}
 \definecolor{RED}{rgb}{1,0,0}
 \definecolor{GREEN}{rgb}{0,1,0}
 \definecolor{BLUE}{rgb}{0,0,1}
 \definecolor{CYAN}{cmyk}{1,0,0,0}
 \definecolor{MAGENTA}{cmyk}{0,1,0,0}
 \definecolor{YELLOW}{cmyk}{0,0,1,0}
}

\makeatother

\usepackage{babel}
\begin{document}

\title{The chiral vortical effect in Wigner function approach}

\author{Jian-Hua Gao}

\affiliation{Shandong Provincial Key Laboratory of Optical Astronomy and Solar-Terrestrial
Environment, Institute of Space Sciences, Shandong University, Weihai,
Shandong 264209, China}

\author{Jin-yi Pang}
\affiliation{Helmholtz-Institut fuer Strahlen- und Kernphysik (Theorie) and Bethe
Center for Theoretical Physics, Universitaet Bonn }

\author{Qun Wang}
\affiliation{Department of Modern Physics, University of Science and Technology
of China, Hefei, Anhui 230026, China}

\begin{abstract}
It is more subtle to obtain the chiral vortical effect (CVE)
than chiral magnetic effect (CME) in quantum transport approach.
To investigate the subtlty of the CVE we present two different derivation
in the Wigner function approach. The first one is based on the method
in our previous work \cite{Gao:2012ix} in which the CVE was derived
under static-equilibrium conditions without details.
We provide a detailed derivation using a more transparent
and powerful method, which can be easily generalized to higher order calculation.
In this derivation of the CVE current, there is an explicit Lorentz covariance.
The second derivation is based on a more general chiral
kinetic theory in a semi-classical expansion of the Wigner function
without assuming static-equilibrium conditions \cite{Gao:2018wmr}.
In this derivation, there is a freedom to choose a reference frame
for the CVE current, so the explicit Lorentz covariance seems to be lost.
Howerver, under static-equilibrium conditions, we show that the CVE current
in this derivation can be decomposed into two parts, identified as the normal
and magnetization current. Each part depends on the reference frame,
but the sum of two parts does give the total CVE current which is independent of the reference frame.
In the comoving frame of the fluid, it can be shown that the normal and
magnetization current give one-third and two-thirds of the total CVE current respectively.
This gives a natural solution to the 'one-third' puzzle in the CVE current
in three-dimensional version of the chiral kinetic theory in the literature.
\end{abstract}

\maketitle

\section{Introduction and summary}

It is well known that rotation and polarization are closely correlated
and can be converted to each other in materials \cite{dehaas:1915,Barnett:1935}.
The same phenomena also exist in high energy heavy ion collisions
(HIC): huge global angular momenta are produced in peripheral collisions
and are expected to induce global polarization of hadrons \cite{Liang:2004ph,Liang:2004xn,Voloshin:2004ha,Betz:2007kg,Becattini:2007sr,Gao:2007bc}.
The global polarization of $\Lambda$ and $\bar{\Lambda}$ hyperons
has been measured by the STAR collaboration \cite{STAR:2017ckg},
which provides a strong evidence for the global rotation in heavy
ion collisions.

The spin-vorticity couplings in statistical-hydro models are widely
used to describe the hadron polarization \cite{Becattini:2013fla,Becattini:2015nva,Becattini:2016gvu,Fang:2016vpj}.
The global polarization of $\Lambda$ and $\bar{\Lambda}$ hyperons
have been calculated through spin-vorticity couplings with vorticity
fields being given by hydrodynamic model simulations \cite{Karpenko:2016jyx,Xie:2017upb}
or transport model simulations \cite{Li:2017slc,Sun:2017xhx,Shi:2017wpk}.
The results are in good agreement with experimental data. For other
model calculations of vorticity fields, see Refs.~\cite{Baznat:2013zx,Csernai:2013bqa,Csernai:2014ywa,Teryaev:2015gxa,Jiang:2016woz,Deng:2016gyh,Ivanov:2017dff}.

The global polarization is an average effect over the whole volume
of the strong interaction matter produced in HIC, so the net vorticity
must be in the direction of the global orbital angular momentum (OAM).
The local vorticity fields contain much more information than the
global one. We assume the following coordiante system for the collisions:
the beams are in the $z$-direction, the OAM is in the $y$-direction,
and the impact parameter is in the $x$-direction. The quadrupole
pattern of $\omega_{y}$ (vorticity in the $y$-direction) in the
reaction plane have been recently studied in hydrodynamic or transport
models \cite{Teryaev:2015gxa,Becattini:2015ska,Jiang:2016woz,Li:2017slc,Shi:2017wpk}.
The similar quadrupole pattern of $\omega_{z}$ in the transverse
plane also exists \cite{Becattini:2017gcx,Karpenko:2017dui,Voloshin:2017kqp}.
Recently a systematic analysis of the quadrupole patterns of $\omega_{x}$,
$\omega_{y}$ and $\omega_{z}$ have been made within the AMPT model
\cite{Xia:2018tes}.

It is still an unsettled question that how such huge global angular
momenta are transferred to the strong interaction matter produced
in HIC, which is under intensive investigation and debate. The question
is profound and highly non-trivial: the spin is a quantum observable
while the vorticity in hydrodynamic description is a classical observable
and how to accomodate the two in a consistent way is not trivial.
Some attempts have been made to include the spin degree of freedom
into relativistic hydrodynamics \cite{Florkowski:2017dyn,Florkowski:2017ruc}.
However there is an ambiguity or freedom on the definition of the
spin tensor out of the total angular momentum one due to the pseudo-gauge
transformation.

In recent years, the Wigner function approach \cite{Heinz:1983nx,Elze:1986qd,Vasak:1987um,Zhuang:1995pd,Florkowski:1995ei,Blaizot:2001nr,Wang:2001dm}
has been revived to describe the chiral magnetic effect (CME) \cite{Vilenkin:1980fu,Kharzeev:2007jp,Fukushima:2008xe,Kharzeev:2015znc}
(for reviews, see, e.g., Ref. \cite{Kharzeev:2013jha,Kharzeev:2015znc,Huang:2015oca})
and chiral vortical effect (CVE) \cite{Vilenkin:1978hb,Erdmenger:2008rm,Banerjee:2008th,Son:2009tf,Gao:2012ix,Hou:2012xg}
for massless fermions \cite{Gao:2012ix,Chen:2012ca,Gao:2015zka,Hidaka:2016yjf,Gao:2017gfq,Gao:2018wmr,Huang:2018wdl}.
The axial vector component of the Wigner function for massive fermions
gives the spin phase-space density \cite{Fang:2016vpj}. Then the
polarization of massive hadrons can be calculated from the Wigner
functions of their constituent quarks \cite{Yang:2017sdk}.

All the proper quantum kinetic approaches should describe both CME and CVE
in a natural and consistent way. However it is more subtle to give rise to the CVE than CME
in quantum transport approach. In this paper we will focus on this issue and try to reveal
the deep structure of the CVE current.
We will present two derivations of the CVE in the Wigner function
based on previous works in Ref. \cite{Gao:2012ix} and Ref. \cite{Gao:2018wmr} by some of us.
In Ref. \cite{Gao:2012ix} on which the first derivation is based
the CVE current was presented but without details.
We will present a detailed derivation of the CVE in a more transparent way
than Ref. \cite{Gao:2012ix}. In Ref.\cite{Gao:2018wmr} on which the second derivation is based,
there was no discussion or derivation of the CVE current in the semi-classical expansion
in the Wigner function formalism. In this paper we will make it up
to discuss the CVE in this approach.

In the first derivation, we will give an improved and detailed derivation of the solutions
to the vector component $\mathscr{J}_{\mu}^{s}(x,p)$ of the Wigner
function for chiral fermions in constant background electromagnetic
fields to the first order in the Planck constant $\hbar$, using the
thermal distribution function under static-equilibrium conditions
including the Killing condition for $\beta^{\mu}=u^{\mu}/T$, where
$u^{\mu}$ and $T$ are the fluid velocity and temperature respectively,
and $s=\pm1$ labels the chirality of chiral fermions. The results
in the improved derivation reproduce the previous ones \cite{Gao:2012ix}
except that the vorticity is replaced by the thermal vorticity times
the temperature. The chiral vortical effect (CVE) is then derived
from the first order solution of $\mathscr{J}_{\mu}^{s}(x,p)$ by
integration over the four-momentum, same as in Ref. \cite{Gao:2012ix}.
The advantage of using the thermal distribution function under static-equilibrium
conditions and constant electromagnetic fields is that they can retain
the explicit Lorentz covariance of the solutions including the CVE
current.

In the second derivation, without assuming any particular form of the distribution
function under static-equilibrium conditions and constant fields,
one can also solve $\mathscr{J}_{\mu}^{s}(x,p)$ by a semiclassical
expansion in powers of $\hbar$ in a systematic way \cite{Gao:2018wmr}.
In this formalism, the spatial components of $\mathscr{J}_{\mu}^{s}(x,p)$
can be derived from the time-component to any order of $\hbar$ \cite{Gao:2018wmr}.
There is a freedom to choose a time-like vector $n^{\mu}$ with $n_{\mu}n^{\mu}=1$
to define the time-component of $\mathscr{J}_{\mu}^{s}(x,p)$ or $n\cdot\mathscr{J}_{\mu}^{s}$.
In other words, there is a freedom to choose a reference frame in
which the time-component of $\mathscr{J}_{\mu}^{s}(x,p)$ is defined
\cite{Gao:2018wmr}. The values of $\mathscr{J}_{\mu}^{s}(x,p)$ should
not depend on $n^{\mu}$. In this formalism we can also obtain the
CVE current as a sum of two parts, which can be identified as the
normal current and magnetization current after using the thermal distribution
under static-equilibrium conditions. Each part depends on $n^{\mu}$,
but the sum is frame independent (or independent of $n^{\mu}$) provided
the distribution function is modified corresponding to the change
of reference frames.

We use the sign convention for the metric tensor $g_{\mu\nu}=\mathrm{diag}(1,-1,-1,-1)$.
We adopt the same sign convention for the fermion charge $Q$ and
$\gamma_{5}$ as in Ref. \cite{Gao:2012ix,Chen:2012ca}. We use $s=\pm1$
to label the chirality of chiral fermions.

\section{Derivation of CVE in the first method}

In this section we will present a detailed and improved derivation of
the CVE based on Ref. \cite{Gao:2012ix}.

\subsection{Wigner function and its solutions in static-equilibrium conditions}

\label{sec:wig}In a background electromagnetic field, the quantum
mechanical analogue of a classical phase-space distribution for fermions
is the gauge invariant Wigner function $W_{\alpha\beta}(x,p)$ which
satisfies the equation of motion \cite{Elze:1986qd,Vasak:1987um},
\begin{equation}
\left(\gamma_{\mu}K^{\mu}-m\right)W(x,p)=0
\end{equation}
where $x=(x_{0},\mathbf{x})$ and $p=(p_{0},\mathbf{p})$ are space-time
and energy-momentum 4-vectors. For the constant field strength $F_{\mu\nu}$,
the operator $K^{\mu}$ is given by $K^{\mu}=p^{\mu}+i\frac{1}{2}\nabla^{\mu}$
with $\nabla^{\mu}=\partial_{x}^{\mu}-QF^{\mu\nu}\partial_{\nu}^{p}$.
The Wigner function can be decomposed in 16 independent generators
of Clifford algebra,
\begin{equation}
W=\frac{1}{4}\left[\mathscr{F}+i\gamma^{5}\mathscr{P}+\gamma^{\mu}\mathscr{V}_{\mu}+\gamma^{5}\gamma^{\mu}\mathscr{A}_{\mu}+\frac{1}{2}\sigma^{\mu\nu}\mathscr{S}_{\mu\nu}\right],\label{eq:wigner-decomp}
\end{equation}
whose coefficients $\mathscr{F}$, $\mathscr{P}$, $\mathscr{V}_{\mu}$,
$\mathscr{A}_{\mu}$ and $\mathscr{S}_{\mu\nu}$ are the scalar, pseudo-scalar,
vector, axial-vector and tensor components of the Wigner function
respectively.

For massless or chiral fermions, the equations for $\mathscr{V}_{\mu}$
and $\mathscr{A}_{\mu}$ are decoupled from other components of the
Wigner function, from which one can obtain independent equations for
vector components $\mathscr{J}_{\mu}^{s}(x,p)$ of the Wigner function
for right-handed $(s=+)$ and left-handed ($s=-$) fermions,
\begin{eqnarray}
p^{\mu}\mathscr{J}_{\mu}^{s}(x,p) & = & 0,\nonumber \\
\nabla^{\mu}\mathscr{J}_{\mu}^{s}(x,p) & = & 0,\nonumber \\
2s(p^{\lambda}\mathscr{J}_{s}^{\rho}-p^{\rho}\mathscr{J}_{s}^{\lambda}) & = & -\hbar\epsilon^{\mu\nu\lambda\rho}\nabla_{\mu}\mathscr{J}_{\nu}^{s},\label{eq:wig-eq}
\end{eqnarray}
where $\mathscr{J}_{\mu}^{s}(x,p)$ are defined as
\begin{eqnarray}
\mathscr{J}_{\mu}^{s}(x,p) & = & \frac{1}{2}[\mathscr{V}_{\mu}(x,p)+s\mathscr{A}_{\mu}(x,p)].
\end{eqnarray}
Note that equations in (\ref{eq:wig-eq}) are valid for constant electromagnetic
field strength.

Under the static-equilibrium condition, one can derive a formal solution
of $\mathscr{J}_{\mu}^{s}$ satisfying Eq. (\ref{eq:wig-eq}) by a
perturbation in powers $\hbar$. Up to $O(\hbar)$, we have the following
solutions to the Wigner functions \cite{Gao:2012ix},
\begin{eqnarray}
\mathscr{J}_{(0)s}^{\rho}(x,p) & = & p^{\rho}f_{s}\delta(p^{2}),\nonumber \\
\mathscr{J}_{(1)s}^{\rho}(x,p) & = & -\frac{s}{2}\tilde{\Omega}^{\rho\alpha}p_{\alpha}\frac{df_{s}}{d(\beta\cdot p)}\delta(p^{2})-\frac{sQ}{p^{2}}\tilde{F}^{\rho\lambda}p_{\lambda}f_{s}\delta(p^{2}).\label{eq:1st-solution}
\end{eqnarray}
The total quantity is given by $\mathscr{J}_{s}^{\rho}=\mathscr{J}_{(0)s}^{\rho}+\hbar\mathscr{J}_{(1)s}^{\rho}$.
In Eq. (\ref{eq:1st-solution}) we have used $\beta^{\mu}=\beta u^{\mu}$
with $\beta=1/T$ being the temperature inverse and with $u^{\mu}$
being the fluid velocity, $\tilde{F}^{\rho\lambda}=\frac{1}{2}\epsilon^{\rho\lambda\mu\nu}F_{\mu\nu}$
denotes the dual of the electromagnetic field strength tensor, $\tilde{\Omega}^{\xi\eta}=\frac{1}{2}\epsilon^{\xi\eta\nu\sigma}\Omega_{\nu\sigma}$
denotes the dual of the thermal vorticity tensor, whose explicit forms
are given in Eq. (\ref{eq:formula}). In Eq. (\ref{eq:1st-solution}),
$f_{s}$ is the distribution function for chiral fermions at the zeroth
order,
\begin{eqnarray}
f_{s}(x,p) & = & \frac{2}{(2\pi)^{3}}\left[\Theta(p_{0})f_{\mathrm{FD}}(\beta\cdot p-\beta\mu_{s})+\Theta(-p_{0})f_{\mathrm{FD}}(-\beta\cdot p+\beta\mu_{s})\right],\label{eq:dist}
\end{eqnarray}
where $p_{0}=u\cdot p$, $f_{\mathrm{FD}}(y)\equiv1/[\exp(y)+1]$
is the Fermi-Dirac distribution function, and $\mu_{s}$ is the chemical
potential for the chirality $s=\pm1$. We can express $\mu_{s}$ in
terms of the scalar and pseudo-scalar (or chiral) chemical potentials,
$\mu_{s}=\mu+s\mu_{5}$.

\subsection{Self-iterative solution of Wigner functions}

\label{sec:derivation}In this section, we will give the derivation
of the first order solution in (\ref{eq:1st-solution}). For simplicity
of notation, we suppress the helicity $s$ from now on. Multiplying
the last line of Eq. (\ref{eq:wig-eq}) by $p_{\lambda}$ and using
the first line of Eq. (\ref{eq:wig-eq}), we obtain
\begin{eqnarray}
2sp^{2}\mathscr{J}^{\rho} & = & -\hbar\epsilon^{\mu\nu\lambda\rho}p_{\lambda}\nabla_{\mu}\mathscr{J}_{\nu},\nonumber \\
 & \rightarrow\nonumber \\
\mathscr{J}^{\rho} & = & \mathcal{J}^{\rho}\delta(p^{2})+\hbar\frac{s}{2p^{2}}\epsilon^{\rho\lambda\mu\nu}p_{\lambda}\nabla_{\mu}\mathscr{J}_{\nu}.\label{eq:j-rho}
\end{eqnarray}
It is easy to see that the second term is satisfied with the first
line of Eq. (\ref{eq:wig-eq}). The first term should also be satisfied
with it, which results in
\begin{eqnarray}
\mathcal{J}^{\rho} & = & p^{\rho}f+\mathscr{X}^{\rho},\label{eq:simple-f}
\end{eqnarray}
with $p\cdot\mathscr{X}=0$. So the last line of Eq. (\ref{eq:j-rho})
becomes
\begin{eqnarray}
\mathscr{J}^{\rho} & = & p^{\rho}f\delta(p^{2})+\mathscr{X}^{\rho}\delta(p^{2})+\hbar\frac{s}{2p^{2}}\epsilon^{\rho\lambda\mu\nu}p_{\lambda}\nabla_{\mu}\mathscr{J}_{\nu}.\label{eq:sol-1}
\end{eqnarray}
It should be pointed out that this expression is updated version compared to the method in Ref.\cite{Gao:2012ix}. The self-iterative feature of this
form reduces the calculation very much and can be more easily generalized to higher order.  Now let us
 insert the above into the last line of Eq. (\ref{eq:wig-eq}),
\begin{eqnarray}
\mathrm{l.h.s.} & = & 2s(p^{\lambda}\mathscr{X}^{\rho}-p^{\rho}\mathscr{X}^{\lambda})\delta(p^{2})+\hbar\frac{1}{p^{2}}p^{[\lambda}\epsilon^{\rho]\delta\mu\nu}p_{\delta}\nabla_{\mu}\mathscr{J}_{\nu},\nonumber \\
\mathrm{r.h.s.} & = & -\hbar\frac{1}{p^{2}}\epsilon^{\lambda\rho\mu\nu}p^{\delta}p_{\delta}\nabla_{\mu}\mathscr{J}_{\nu}.
\end{eqnarray}
Then the last line of Eq. (\ref{eq:wig-eq}) becomes
\begin{eqnarray}
2s(p^{\lambda}\mathscr{X}^{\rho}-p^{\rho}\mathscr{X}^{\lambda})\delta(p^{2}) & = & \hbar\frac{1}{p^{2}}[p^{\mu}\epsilon^{\nu\delta\lambda\rho}+p^{\nu}\epsilon^{\delta\lambda\rho\mu}]p_{\delta}\nabla_{\mu}\mathscr{J}_{\nu}\nonumber \\
 & = & \hbar\frac{1}{p^{2}}[\nabla_{\mu}\mathscr{J}_{\nu}-\nabla_{\nu}\mathscr{J}_{\mu}]\epsilon^{\nu\delta\lambda\rho}p^{\mu}p_{\delta},\label{eq:cke-x}
\end{eqnarray}
where we have used
\begin{equation}
\epsilon^{\lambda\rho\mu\nu}p^{\delta}+\epsilon^{\rho\mu\nu\delta}p^{\lambda}+\epsilon^{\mu\nu\delta\lambda}p^{\rho}+\epsilon^{\nu\delta\lambda\rho}p^{\mu}+\epsilon^{\delta\lambda\rho\mu}p^{\nu}=0.
\end{equation}
We see from Eq. (\ref{eq:cke-x}) that $\mathscr{X}^{\rho}$ is at
least of $O(\hbar)$. Now we can rewrite the last line of Eq. (\ref{eq:wig-eq})
to this form
\begin{eqnarray}
\hbar\nabla_{[\sigma}\mathscr{J}_{\xi]} & = & 2s\epsilon_{\sigma\xi\lambda\rho}p^{\lambda}\mathscr{J}^{\rho}.\label{eq:cke-2}
\end{eqnarray}
Using Eq. (\ref{eq:cke-2}) in the last line of Eq. (\ref{eq:cke-x}),
without the $1/p^{2}$ factor, we obtain
\begin{eqnarray}
\hbar\nabla_{[\mu}\mathscr{J}_{\nu]}\epsilon^{\nu\delta\lambda\rho}p^{\mu}p_{\delta} & = & 2s\epsilon_{\mu\nu\alpha\beta}\epsilon^{\nu\delta\lambda\rho}p^{\alpha}p^{\mu}p_{\delta}\mathscr{J}^{\beta}=0.\label{eq:nabla-mu-nu}
\end{eqnarray}
Note that $p^{\mu}\nabla_{[\mu}\mathscr{J}_{\nu]}$ contains a term
$\sim p^{2}\delta(p^{2})$ which is vanishing. But when multiplying
Eq. (\ref{eq:nabla-mu-nu}) by a prefactor $1/p^{2}$, i.e. as in
the the last line of Eq. (\ref{eq:cke-x}), the term $p^{2}\delta(p^{2})$
in $p^{\mu}\nabla_{[\mu}\mathscr{J}_{\nu]}$ gives non-vanishing value.
Now we try to extract the $p^{2}\delta(p^{2})$ term in $p^{\mu}\nabla_{[\mu}\mathscr{J}_{\nu]}$
by using Eq. (\ref{eq:sol-1}),
\begin{eqnarray}
p^{\mu}\nabla_{[\mu}\mathscr{J}_{\nu]} & = & p^{\mu}\nabla_{[\mu}[p_{\nu]}f\delta(p^{2})]+p^{\mu}\nabla_{[\mu}[\mathscr{X}_{\nu]}\delta(p^{2})]\nonumber \\
 &  & +\hbar\frac{s}{2p^{2}}p^{\mu}\epsilon_{[\nu\alpha\beta\gamma}\nabla_{\mu]}[p^{\alpha}\nabla^{\beta}\mathscr{J}^{\gamma}].\label{eq:nabla-mu-nu-1}
\end{eqnarray}
We look at the first term $p^{\mu}\nabla_{[\mu}[p_{\nu]}f\delta(p^{2})]$,
\begin{eqnarray}
p^{\mu}\nabla_{\mu}[p_{\nu}f\delta(p^{2})] & = & p^{\mu}p_{\nu}\delta(p^{2})\nabla_{\mu}f-Qp^{\mu}F_{\mu\nu}f\delta(p^{2}),\nonumber \\
p^{\mu}\nabla_{\nu}[p_{\mu}f\delta(p^{2})] & = & p^{2}\delta(p^{2})\nabla_{\nu}f-Qp^{\mu}F_{\nu\mu}f\delta(p^{2})-2Qp^{2}F_{\nu\rho}p^{\rho}f\delta^{\prime}(p^{2}).\label{eq:nabla-j0}
\end{eqnarray}
So the last line of Eq. (\ref{eq:cke-x}) has a non-zero contribution
\begin{eqnarray}
\hbar\frac{1}{p^{2}}\nabla_{[\mu}\mathscr{J}_{\nu]}\epsilon^{\nu\delta\lambda\rho}p^{\mu}p_{\delta} & \rightarrow & -\hbar\frac{1}{p^{2}}[p^{2}\delta(p^{2})\nabla_{\nu}f]\epsilon^{\nu\delta\lambda\rho}p_{\delta}\nonumber \\
 & = & -\hbar\delta(p^{2})\epsilon^{\nu\delta\lambda\rho}p_{\delta}\nabla_{\nu}f.\label{eq:non-van}
\end{eqnarray}
The last two terms of Eq. (\ref{eq:nabla-mu-nu-1}) are of higher
order and will not be considered at the first order.

From Eqs. (\ref{eq:cke-x},\ref{eq:non-van}), we finally obtain
\begin{eqnarray}
(p^{\lambda}\mathscr{X}^{\rho}-p^{\rho}\mathscr{X}^{\lambda})\delta(p^{2}) & = & \hbar\frac{s}{2}\delta(p^{2})\epsilon^{\delta\nu\lambda\rho}p_{\delta}\nabla_{\nu}f.\label{eq:x-f}
\end{eqnarray}
We can multiply a factor $\epsilon_{\mu\alpha\lambda\rho}$ and sum
over $\lambda\rho$,
\begin{eqnarray}
2\epsilon_{\mu\alpha\lambda\rho}p^{\lambda}\mathscr{X}^{\rho}\delta(p^{2}) & = & -\hbar s\delta(p^{2})(p_{\mu}\nabla_{\alpha}f-p_{\alpha}\nabla_{\mu}f).\label{eq:sol-2}
\end{eqnarray}
The distribution for the right-handed or left-handed fermions is given
by Eq. (\ref{eq:dist}). We will use following notations: $\beta^{\rho}=\beta u^{\rho}$
and $\bar{\mu}_{s}=\beta\mu_{s}$ (again, in the following we suppress
the helicity label $s$ in the chemical potential). Let's calculate
$\nabla_{\nu}f$,
\begin{eqnarray}
\nabla_{\nu}f & = & \frac{\partial f}{\partial(\beta\cdot p)}\left[p^{\sigma}\frac{\partial\beta_{\sigma}}{\partial x^{\nu}}-\frac{\partial\bar{\mu}}{\partial x^{\nu}}-\beta QE_{\nu}\right]\nonumber \\
 & = & f^{\prime}\left[\frac{1}{2}p^{\sigma}(\partial_{\nu}\beta_{\sigma}-\partial_{\sigma}\beta_{\nu})+\frac{1}{2}p^{\sigma}(\partial_{\nu}\beta_{\sigma}+\partial_{\sigma}\beta_{\nu})-\partial_{\nu}\bar{\mu}-\beta QE_{\nu}\right]\nonumber \\
 & \rightarrow & \frac{1}{2}f^{\prime}p^{\sigma}(\partial_{\nu}\beta_{\sigma}-\partial_{\sigma}\beta_{\nu})=f^{\prime}\Omega_{\nu\sigma}p^{\sigma},\label{eq:nabla-f}
\end{eqnarray}
where we have used the shorthand notation $f^{\prime}\equiv\frac{\partial f}{\partial(\beta\cdot p)}$,
then $\frac{\partial f_{s}}{\partial\bar{\mu}_{s}}=-f^{\prime}$ and
$\Omega_{\nu\sigma}=\frac{1}{2}(\partial_{\nu}\beta_{\sigma}-\partial_{\sigma}\beta_{\nu})$.
Most importantly we have used the following static-equilbrium conditions
to reach the last line of Eq. (\ref{eq:nabla-f}),
\begin{eqnarray}
 &  & \partial_{\alpha}\beta_{\sigma}+\partial_{\sigma}\beta_{\alpha}=0,\nonumber \\
 &  & \partial_{\alpha}\bar{\mu}_{s}=-\beta QE_{\alpha},\label{eq:killing-condition}
\end{eqnarray}
where the first line of Eq. (\ref{eq:killing-condition}) is the Killing
condition for $\beta^{\mu}$. The Killing condition leads to the solution
$\beta^{\mu}=b^{\mu}-\Omega^{\mu\nu}x_{\nu}$ with $b^{\mu}$ and
$\Omega^{\mu\nu}$ being constants. The second line of Eq. (\ref{eq:killing-condition})
leads to equations for the fermion number and chiral chemical potential,
$\partial_{\lambda}\bar{\mu}=-\beta QE_{\lambda}$ and $\partial_{\nu}\bar{\mu}_{5}=0$.
It also leads to the integrability condition for constant field strength,
\begin{eqnarray}
F_{\mu}^{\;\rho}\partial_{\nu}\beta_{\rho}-F_{\nu}^{\;\rho}\partial_{\mu}\beta_{\rho} & = & 0,
\end{eqnarray}
or in a compact form
\begin{eqnarray}
F_{\mu}^{\;\rho}\Omega_{\nu\rho}-F_{\nu}^{\;\rho}\Omega_{\mu\rho} & = & 0.
\end{eqnarray}
Then we can further derive the following identities
\begin{eqnarray}
\frac{1}{2}\tilde{\Omega}^{\mu\alpha}F_{\mu\alpha}p^{2} & = & \Omega^{\mu\alpha}\tilde{F}_{\mu\rho}p_{\alpha}p^{\rho}+\tilde{\Omega}^{\mu\alpha}F_{\mu\rho}p_{\alpha}p^{\rho}\nonumber \\
\tilde{F}^{\mu\alpha}F_{\mu\sigma}p_{\alpha}p^{\sigma} & = & \frac{1}{4}p^{2}\tilde{F}^{\rho\lambda}F_{\rho\lambda}\nonumber \\
\tilde{\Omega}^{\mu\alpha}\Omega_{\mu\sigma}p_{\alpha}p^{\sigma} & = & \frac{1}{4}p^{2}\tilde{\Omega}^{\rho\lambda}\Omega_{\rho\lambda}.\label{eq:int-condition}
\end{eqnarray}
The proof of Eq. (\ref{eq:int-condition}) is given in Eqs. (\ref{eq:first-id},\ref{eq:second-id}).

So Eq. (\ref{eq:sol-2}) can be simplified as (we suppress $\delta(p^{2})$)
\begin{eqnarray}
2\epsilon_{\mu\alpha\lambda\rho}p^{\lambda}\mathscr{X}^{\rho} & = & -\hbar s(p_{\mu}\Omega_{\alpha\sigma}p^{\sigma}-p_{\alpha}\Omega_{\mu\sigma}p^{\sigma})f^{\prime},
\end{eqnarray}
or put in another form by contraction with $\epsilon^{\mu\alpha\gamma\delta}$,
\begin{eqnarray}
p^{\mu}\mathscr{X}^{\nu}-p^{\nu}\mathscr{X}^{\mu} & = & \hbar\frac{s}{2}\epsilon^{\mu\nu\alpha\gamma}p_{\alpha}\Omega_{\gamma\sigma}p^{\sigma}f^{\prime}\nonumber \\
 & = & -\hbar\frac{s}{4}\epsilon^{\mu\nu\alpha\gamma}\epsilon_{\gamma\sigma\rho\xi}\tilde{\Omega}^{\rho\xi}p_{\alpha}p^{\sigma}f^{\prime}\nonumber \\
 & = & -\hbar\frac{s}{4}\delta_{\sigma\rho\xi}^{\mu\nu\alpha}\tilde{\Omega}^{\rho\xi}p_{\alpha}p^{\sigma}f^{\prime}\nonumber \\
 & = & -\hbar\frac{s}{2}(p^{\mu}p_{\rho}\tilde{\Omega}^{\nu\rho}-p^{\nu}p_{\rho}\tilde{\Omega}^{\mu\rho})f^{\prime},\label{eq:sol-3}
\end{eqnarray}
where we have dropped the $p^{2}\tilde{\Omega}^{\mu\nu}$ term inside
the brackets due to $\delta(p^{2})$, $\delta_{\sigma\rho\xi}^{\mu\nu\alpha}\equiv-\epsilon^{\gamma\mu\nu\alpha}\epsilon_{\gamma\sigma\rho\xi}$
is given by Eq. (\ref{eq:epsilon}), and $\delta_{\sigma\rho\xi}^{\mu\nu\alpha}\tilde{\Omega}^{\rho\xi}p_{\alpha}p^{\sigma}$
is given by Eq. (\ref{eq:del-o-pp}). From the last line of Eq. (\ref{eq:sol-3})
we obtain
\begin{equation}
\mathscr{X}^{\mu}=-\hbar\frac{s}{2}p_{\rho}\tilde{\Omega}^{\mu\rho}f^{\prime}.\label{eq:x-mu}
\end{equation}
Then Eq. (\ref{eq:sol-1}) is in the form
\begin{equation}
\mathscr{J}^{\rho}=p^{\rho}f\delta(p^{2})-\hbar\frac{s}{2}p_{\sigma}\tilde{\Omega}^{\rho\sigma}f^{\prime}\delta(p^{2})+\hbar\frac{s}{2p^{2}}\epsilon^{\rho\lambda\mu\nu}p_{\lambda}\nabla_{\mu}\mathscr{J}_{\nu},\label{eq:sol-4}
\end{equation}
where $f^{\prime}\equiv\frac{\partial f}{\partial(\beta\cdot p)}$.
The first term is the zeroth order contribution (generally it can
include higher order contributions, but here we neglect this possibility
for simplicity), while the second and third term are at least the
first order contribution. We can replace $\mathscr{J}_{\nu}$ in Eq.
(\ref{eq:sol-4}) with the zeroth order contribution $p_{\nu}f\delta(p^{2})$
to obtain the zeroth and first order contribution given in Eq. (\ref{eq:1st-solution}).

Now we check if the solutions (\ref{eq:1st-solution}) satisfy the
second line of Eq. (\ref{eq:wig-eq}). Let us look at the zeroth order
solution
\begin{eqnarray}
\nabla_{\mu}\mathscr{J}_{(0)}^{\mu} & = & \nabla_{\mu}[p^{\mu}f\delta(p^{2})]\nonumber \\
 & = & \delta(p^{2})p^{\mu}\nabla_{\mu}f=0,
\end{eqnarray}
where we have used $\nabla_{\mu}f=f^{\prime}\Omega_{\mu\sigma}p^{\sigma}$
in Eq. (\ref{eq:nabla-f}). For the first order solution in (\ref{eq:1st-solution}),
following Eqs. (\ref{eq:nabla-j-om},\ref{eq:nabla-j-em}), so we
can verify
\begin{equation}
\nabla_{\mu}\mathscr{J}_{(1)}^{\mu}=0
\end{equation}
hold for constant vortcity and field strength tensor. So we see that
the second line of Eq. (\ref{eq:wig-eq}) does hold. From Eqs. (\ref{eq:nabla-j-om},\ref{eq:nabla-j-em}),
we see that if there is an electromagnetic field in a system of charged
fermions as shown in the electromagnetic term in $\mathscr{J}_{(1)}^{\mu}$,
it will make fermions rotate which results in a vorticity term in
$\mathscr{J}_{(1)}^{\mu}$. Therefore both terms coexist in the first
order solutions (\ref{eq:1st-solution}). This is consistent with
the observation that both the vorticity and magnetic field terms in
a system of charged fermions must coexist for the second law of thermodynamics
to be satisfied \cite{Son:2009tf,Pu:2010as,Ozonder:2010zy}.

\subsection{CVE and CME current }

\label{sec:cve-cme-current}The current can be obtained from $\mathscr{J}_{s}^{\rho}$
in Eq. (\ref{eq:1st-solution}) by inetgration over $p$,
\begin{eqnarray}
j_{s}^{\rho} & = & \int d^{4}p\mathscr{J}_{s}^{\rho}.
\end{eqnarray}
In this section we recover the chirality label $s$. We can use the
thermal distribution (\ref{eq:dist}) to evaluate $j_{s}^{\rho}$
in equilibrium. The zeroth order contribution is given by $\mathscr{J}_{(0)s}^{\rho}$,
\begin{eqnarray}
j_{(0)s}^{\rho} & = & \int d^{4}pp^{\rho}f_{s}\delta(p^{2})=u^{\rho}\int d^{4}pp_{0}f_{s}\delta(p^{2})\nonumber \\
 & = & u^{\rho}\int\frac{d^{3}p}{(2\pi)^{3}}\left(f_{\mathrm{FD}}^{+}-f_{\mathrm{FD}}^{-}\right)\nonumber \\
 & = & N_{s}u^{\rho},\label{eq:j0}
\end{eqnarray}
where we have used the decomposition $p^{\rho}=p_{0}u^{\rho}+\bar{p}^{\rho}$
with $p_{0}\equiv u\cdot p$ and $\bar{p}\cdot u=0$, we have used
the shorthand notation $f_{\mathrm{FD}}^{\pm}=f_{\mathrm{FD}}(\beta E_{p}\mp\beta\mu_{s})$
with $E_{p}=\sqrt{-\bar{p}_{\mu}\bar{p}^{\mu}}$, and $N_{s}$ is
the fermion number density of chiral fermions. In Eq. (\ref{eq:j0}),
we have dropped the term proportional to the spatial part $\bar{p}^{\rho}$
since its integral is zero for the thermal distribution. We can obtain
the CVE current from the $\tilde{\Omega}^{\rho\sigma}$ term in Eq.
(\ref{eq:1st-solution}),
\begin{eqnarray}
j_{s,\omega}^{\rho} & = & -\hbar\frac{s}{2}\tilde{\Omega}^{\rho\sigma}\int d^{4}pp_{\sigma}\frac{\partial f_{s}}{\partial(\beta\cdot p)}\delta(p^{2})\nonumber \\
 & = & \hbar s\omega^{\rho}\frac{1}{2}\int\frac{d^{3}p}{(2\pi)^{3}}\left[f_{\mathrm{FD}}^{+}(1-f_{\mathrm{FD}}^{+})+f_{\mathrm{FD}}^{-}(1-f_{\mathrm{FD}}^{-})\right]\nonumber \\
 & = & \omega^{\rho}\frac{s\hbar}{2\pi^{2}\beta}\int_{0}^{\infty}dE_{p}E_{p}\left(f_{\mathrm{FD}}^{+}+f_{\mathrm{FD}}^{-}\right)\nonumber \\
 & = & T\xi_{s}\omega^{\rho},\label{eq:current-omega}
\end{eqnarray}
where we have the contribution from $\bar{p}_{\sigma}$ since its
integral is zero for the thermal distribution. In Eq. (\ref{eq:current-omega})
$\omega^{\rho}$ is the thermal vorticity and $\xi_{s}$ is the CVE
coefficient for the chiral fermion with the chirality $s$. The vector
and axial vector currents in the chiral vortical effect are given
by,
\begin{eqnarray}
j^{\mu}(\omega) & = & (\xi_{+}+\xi_{-})\omega^{\mu}=\xi\omega^{\mu},\nonumber \\
j_{5}^{\mu}(\omega) & = & (\xi_{+}-\xi_{-})\omega^{\mu}=\xi_{5}\omega^{\mu},\label{eq:cve-curr}
\end{eqnarray}
where $\xi=\mu\mu_{5}/\pi^{2}$ and $\xi_{5}=T^{2}/6+(\mu^{2}+\mu_{5}^{2})/(2\pi^{2})$
are coefficients in Eqs. (22-23) of Ref. \cite{Gao:2012ix}.

It is easy to check that the $F_{\mu\nu}$ term in Eq. (\ref{eq:1st-solution})
gives the CME current in equilibrium,
\begin{eqnarray}
j_{s,B}^{\rho} & = & \hbar sQ\tilde{F}^{\rho\lambda}\int d^{4}pp_{\lambda}\delta^{\prime}(p^{2})f_{s}\nonumber \\
 & = & \hbar sQ\beta B^{\rho}\frac{1}{2}\int\frac{d^{3}p}{(2\pi)^{3}}\frac{1}{E_{p}}\left[f_{\mathrm{FD}}^{+}(1-f_{\mathrm{FD}}^{+})-f_{\mathrm{FD}}^{-}(1-f_{\mathrm{FD}}^{-})\right]\nonumber \\
 & = & B^{\rho}\frac{sQ\hbar}{4\pi^{2}}\int_{0}^{\infty}dE_{p}\left(f_{\mathrm{FD}}^{+}-f_{\mathrm{FD}}^{-}\right)\nonumber \\
 & = & \xi_{B}^{s}B^{\rho},\label{eq:cme}
\end{eqnarray}
where we have dropped the term with spatial momentum $\bar{p}_{\lambda}$
and the surface term in $p_{0}$ in the third line. Note that $\xi_{B}^{s}$
is the CME conductivity for chiral fermions with the chirality $s$.
Then one can reproduce the chiral magnetic effect for the vector and
axial vector currents as
\begin{eqnarray}
j_{B}^{\mu} & = & (\xi_{B}^{+}+\xi_{B}^{-})B^{\mu}=\xi_{B}B^{\mu},\nonumber \\
j_{5,B}^{\mu} & = & (\xi_{B}^{+}-\xi_{B}^{-})B^{\mu}=\xi_{B5}B^{\mu},\label{eq:cme-current}
\end{eqnarray}
where $\xi_{B}=Q\mu_{5}/(2\pi^{2})$ and $\xi_{B5}=Q\mu/(2\pi^{2})$
are coefficients in Eqs. (22-23) of Ref. \cite{Gao:2012ix}.

\section{Derivation of CVE in the second method}

In Sect. \ref{sec:wig}-\ref{sec:cve-cme-current}, we have derived
and applied the first order solutions of the Wigner function with
the assumption of the thermal distribution function under static-equilibrium
conditions and constant electromagnetic fields. The advantage of this
method is that the explicit Lorentz covariance of the solutions is
retained. In this section, we will present the second derivation of CME and CVE
based on Ref. \cite{Gao:2018wmr}.

Without assuming the form of the distribution function under static-equilibrium
conditions and constant electromagnetic fields, one can still solve
the Wigner function $\mathscr{J}^{\mu}$ in the framework of semiclassical
expansion \cite{Gao:2018wmr}. To any order of $\hbar$ that it has
been shown that only one component of $\mathscr{J}^{\mu}$ is independent
while other three components can be derived from the independent one.
One can choose, for example, the time-component $\mathscr{J}^{0}$
as the independent one which defines the distribution function. But
there is a freedom to choose any reference frame to define the time-component,
so all spatial components which are orthogonal to it can then be derived
in this reference frame. This superficially break the Lorentz covariance
\cite{Chen:2014cla,Duval:2014cfa,Stone:2014fja,Stone:2015kla,Huang:2018aly}
of $\mathscr{J}^{\mu}$. The requirement that $\mathscr{J}_{\mu}$
should not depend on the choice of the reference frame in which $\mathscr{J}^{0}$
is defined leads to the corresponding change of the distribution function
\cite{Gao:2018wmr}. In this method we can also derive the zeroth
and first order soultion $\mathscr{J}_{(0,1)}^{\mu}$. The CVE current
can be obtained from $\mathscr{J}_{(1)}^{\mu}$ by momentum integration.

The reference frame is characterized by a time-like vector $n^{\mu}$
with normalization $n^{2}=1$. In the following, we will assume $n^{\mu}(x)$
is a general vector with space-time dependence.

In this section we will show that the CVE current from $\mathscr{J}_{(1)}^{\mu}$
obtained in the semiclassical expansion in Ref. \cite{Gao:2018wmr}
has two parts, which can be identified as the normal current and magnetization
current after using the thermal distribution and static-equilibrium
conditions. Each part, the normal or magnetization current, depends
on $n^{\mu}$, but the sum of two is frame independent (or independent
of $n^{\mu}$) provided the distribution function is changed in a
given manner corresponding to the change of reference frames.

In  Ref. \cite{Gao:2018wmr},
the vector component of the Wigner function for chiral fermions up to $O(\hbar)$ is given by ,
\begin{eqnarray}
\mathscr{J}_{(0)}^{\mu} & = & p^{\mu}f_{(0)}\delta(p^{2}),\nonumber \\
\mathscr{J}_{(1)}^{\mu} & = & p^{\mu}f_{(1)}\delta(p^{2})
+ p^\mu \frac{s Q  }{n\cdot p}n_\nu \tilde F^{\nu\lambda} p_\lambda f_{(0)}  \delta'\left(p^2\right)
-\frac{s }{2n\cdot p}\epsilon^{\mu\nu\rho\sigma} n _\nu   \nabla_\sigma
\left[p_\rho  f_{(0)} \delta\left(p^2\right)\right],\label{Jmu-1}
\end{eqnarray}

where $f_{(0,1)}$ are arbitrary functions of space and time where
the indices '(0)' and '(1)' label the orders in $\hbar$. With non-constant
$n^{\mu}(x)$ the derivation of Eq. (\ref{Jmu-1}) is given in Appendix
\ref{sec:app-2}. Obviously the second term of $\mathscr{J}_{(1)}^{\mu}$
in Eq. (\ref{Jmu-1}) depends on $n^{\mu}$. We can extract $f_{(0,1)}$
by $f_{(0,1)}\delta(p^2)=(n\cdot\mathscr{J}_{(0,1)})/(n\cdot p)$. We see that
the form of $\mathscr{J}_{(1)}^{\mu}$ in Eq. (\ref{Jmu-1}) is different
from Eq. (\ref{eq:1st-solution}) derived based on the thermal distribution
function under static-equilibrium conditions.

From the definition $f_{(0,1)}\delta(p^2)=(n\cdot\mathscr{J}_{(0,1)})/(n\cdot p)$
and Eq. (\ref{eq:wig-eq}) at the zeroth and first order we obtain
the variations of distribution functions \cite{Gao:2018wmr}
\begin{eqnarray}
\delta f_{(0)} \delta(p^2)& = & 0,\nonumber \\
\delta f_{(1)} \delta(p^2) & = &
-sQ\left(\frac{n'_\nu \tilde F^{\nu\lambda}p_\lambda}{n'\cdot p}
-\frac{n_\nu \tilde F^{\nu\lambda}p_\lambda}{n\cdot p}\right)\delta'\left(p^2\right) f_{(0)}
 -s\frac{\epsilon^{\lambda\nu\rho\sigma}n_{\lambda}n_{\nu}^{\prime}}
 {2\left(n^{\prime}\cdot p\right)\left(n\cdot p\right)}\nabla_{\rho}\left[p_{\sigma} f_{(0)}\delta\left(p^2\right) \right].\label{eq:f0-f1-var}
\end{eqnarray}

Note that $\delta f_{(1)}$ in Eq. (\ref{eq:f0-f1-var}) is called
side-jump \cite{Chen:2014cla} whose derivation follows Eq. (\ref{eq:wig-eq})
for the zeroth and first order terms. With Eq. (\ref{eq:wig-eq})
and (\ref{Jmu-1}), we can also verify that the variation of the second and third
terms in Eq. (\ref{Jmu-1}) does give $-p^{\mu}\delta f_{(1)}\delta(p^2)$
\begin{eqnarray}
\delta\mathscr{J}_{(1)}^{\mu}(2) & = &-s Q p^\mu \left(\frac{n'_\nu \tilde F^{\nu\lambda}p_\lambda}{n'\cdot p}
-\frac{n_\nu \tilde F^{\nu\lambda}p_\lambda}{n\cdot p}\right)\delta'\left(p^2\right) f_{(0)}
 -s\frac{\left(n\cdot p\right)n_{\nu}^{\prime}-\left(n^{\prime}\cdot p\right)n_{\nu}}{2\left(n^{\prime}\cdot p\right)(n\cdot p)}
 \epsilon^{\mu\nu\rho\sigma}\nabla_{\sigma}\left[ p_{\rho} f_{(0)}\delta(p^{2})\right]\nonumber \\
 & = & -p^{\mu}\left(\frac{n^{\prime}\cdot\mathscr{J}_{(1)}}{n^{\prime}\cdot p}-\frac{n\cdot\mathscr{J}_{(1)}}{n\cdot p}\right)
 =-p^{\mu}\delta f_{(1)}\delta(p^2).
\end{eqnarray}
So we conclude that the vector component does not change as it should
be when one changes the reference frame $n^{\mu}\rightarrow n^{\prime\mu}$,

\begin{equation}
\delta\mathscr{J}_{(0,1)}^{\mu}=\mathscr{J}_{(0,1)}^{\prime\mu}-\mathscr{J}_{(0,1)}^{\mu}=0\label{eq:d-j-lorentz}
\end{equation}

Since $\delta f_{(0)}=0$, we can assume that the scalar function
$f_{(0)}$ only depends on $\beta\cdot p$ and $\mu_s$ where $\beta^{\mu}=u^{\mu}/T$.
If we further assume $\partial_{\alpha}\bar{\mu}_{s}=-\beta QE_{\alpha}$ and
the Killing condition $\partial_{\mu}^{x}\beta_{\nu}+\partial_{\nu}^{x}\beta_{\mu}=0$
in Eq. (\ref{eq:killing-condition}) which implies a constant thermal
vorticity tensor $\Omega_{\mu\nu}=(\partial_{\mu}^{x}\beta_{\nu}-\partial_{\nu}^{x}\beta_{\mu})/2$,
it is easy to show that $f_{(0)}$ satisfies the chiral kinetic equation
$\delta(p^{2})p^{\mu}\partial_{\mu}^{x}f_{(0)}=0$. From the form
of $\delta f_{(1)}$ in Eq. (\ref{eq:f0-f1-var}) we then obtain
\begin{eqnarray}
\delta f_{(1)} & = & -s\frac{p^{\mu}\epsilon^{\lambda\nu\rho\sigma}n_{\lambda}n_{\nu}^{\prime}p_{\sigma}\partial_{\rho}^{x}\beta_{\mu}}{2\left(n^{\prime}\cdot p\right)\left(n\cdot p\right)}\frac{df_{(0)}}{d(\beta\cdot p)}\nonumber \\
 & = & -s\frac{p^{\mu}\epsilon^{\lambda\nu\rho\sigma}n_{\lambda}n_{\nu}^{\prime}p_{\sigma}\Omega_{\rho\mu}}{2\left(n^{\prime}\cdot p\right)\left(n\cdot p\right)}\frac{df_{(0)}}{d(\beta\cdot p)}\nonumber \\
 & = & -s\frac{n_{\alpha}^{\prime}p_{\gamma}\tilde{\Omega}^{\alpha\gamma}}{2\left(n^{\prime}\cdot p\right)}\frac{df_{(0)}}{d(\beta\cdot p)}+s\frac{n_{\alpha}p_{\gamma}\tilde{\Omega}^{\alpha\gamma}}{2\left(n\cdot p\right)}\frac{df_{(0)}}{d(\beta\cdot p)},
\end{eqnarray}
where we have used Eqs. (\ref{eq:formula},\ref{eq:epsilon}). Note that only vorticity terms contribute and electromagnetic field does not contribute.
 Now let us express $f_{(1)}$ as
\begin{eqnarray}
f_{(1)} & = & \tilde{f}_{(1)}-s\frac{n_{\alpha}p_{\gamma}\tilde{\Omega}^{\alpha\gamma}}{2\left(n\cdot p\right)}\frac{df_{(0)}}{d(\beta\cdot p)},\label{f1}
\end{eqnarray}
where $\tilde{f}_{(1)}$ does not depend on the reference vector $n^{\mu}$,
i.e. $\delta\tilde{f}_{(1)}=0$. We can choose the specific solution
with $\tilde{f}_{(1)}=0$. Substituting Eq. (\ref{f1}) into Eq.(\ref{Jmu-1})
yields
\begin{eqnarray}
\mathscr{J}_{(1)}^{\mu} & = & \mathscr{J}_{(1)B}^{\mu} + \mathscr{J}_{(1)\omega}^{\mu}
\end{eqnarray}
where
\begin{eqnarray}
\mathscr{J}_{(1)B}^{\mu} & = & \frac{s Q  }{n\cdot p}p^\mu n_\nu \tilde F^{\nu\lambda} p_\lambda f_{(0)}  \delta'\left(p^2\right)
+\frac{s Q}{2n\cdot p}\epsilon^{\mu\nu\rho\sigma} n _\nu   F_{\sigma\lambda} f_{(0)} \partial_p^\lambda
\left[p_\rho   \delta\left(p^2\right)\right] \nonumber\\
&=& sQ \tilde F^{\mu\lambda}p_\lambda f_{(0)} \delta'(p^2),\label{JmuB-1}\\
\mathscr{J}_{(1)\omega}^{\mu} & = & -p^{\mu}\frac{s}{2\left(n\cdot p\right)}n_{\alpha}p_{\gamma}\tilde{\Omega}^{\alpha\gamma}\frac{df_{(0)}}{d(\beta\cdot p)}\delta(p^{2})
-\frac{s}{2n\cdot p}p^\lambda \epsilon^{\mu\nu\rho\sigma}n_{\nu}p_{\rho} \Omega_{\sigma\lambda }\frac{df_{(0)}}{d(\beta\cdot p)} \delta(p^{2})\nonumber \\
& = & -\frac{s}{2}\tilde{\Omega}^{\mu\nu}p_{\nu}\frac{df_{(0)}}{d(\beta\cdot p)}\delta(p^{2}),\label{eq:j-mu-1-inv}
\end{eqnarray}
where we used Eqs. (\ref{eq:formula},\ref{eq:epsilon}) to obtain
the last equalities in Eqs.(\ref{JmuB-1},\ref{eq:j-mu-1-inv}). We see that $\mathscr{J}_{(1)B}^{\mu}$ and $\mathscr{J}_{(1)\omega}^{\mu}$ are both
 independent of $n^{\mu}$ and matches the first term of Eq. (\ref{eq:1st-solution}) by setting $f_{(0)}=f_{s}$.

Now we evaluate the contributions to the CVE current in (\ref{eq:current-omega})
from the two terms of the first equality in Eq. (\ref{eq:j-mu-1-inv})
which we call $j_{s,\omega}^{\mu}(1)$ and $j_{s,\omega}^{\mu}(2)$,
\begin{eqnarray}
j_{s,\omega}^{\mu}(1) & = & -\hbar\frac{s}{2}u_{\alpha}\tilde{\Omega}^{\alpha\gamma}\int d^{4}p\frac{1}{(u\cdot p)}p^{\mu}p_{\gamma}\frac{df_{s}}{d(\beta\cdot p)}\delta(p^{2})\nonumber \\
 & = & -\hbar\frac{s}{2}\omega^{\gamma}\int\frac{d^{3}p}{(2\pi)^{3}}\frac{\bar{p}^{\mu}\bar{p}_{\gamma}}{E_{p}^{2}}\left[f_{\mathrm{FD}}^{+}(1-f_{\mathrm{FD}}^{+})+f_{\mathrm{FD}}^{-}(1-f_{\mathrm{FD}}^{-})\right]\nonumber \\
 & = & \frac{1}{3}T\xi_{s}\omega^{\mu},\nonumber \\
j_{s,\omega}^{\mu}(2) & = & -\hbar\frac{s}{2}\epsilon^{\mu\nu\rho\sigma}u_{\nu}\int d^{4}p\frac{1}{u\cdot p}p_{\rho}(\partial_{\sigma}^{x}f_{s})\delta(p^{2})\nonumber \\
 & = & \frac{2}{3}T\xi_{s}\omega^{\mu},\label{eq:j1-j2}
\end{eqnarray}
where we have assumed $n^{\mu}=u^{\mu}$ and take $f_{(0)}=f_{s}(x,p)$
given by Eq. (\ref{eq:dist}). Note that $\Omega^{\mu\nu}$ and $\tilde{\Omega}^{\mu\nu}$
are thermal vorticity tensors and $\omega^{\mu}$ is the thermal vorticity
vector which are all dimensionless. We see in Eq. (\ref{eq:j1-j2})
that $j_{s,\omega}^{\mu}(1)$ and $j_{s,\omega}^{\mu}(2)$ contribute
to the full CVE current by 1/3 and 2/3 repectively. In order to see
the physical meaning of $j_{s,\omega}^{\mu}(1)$ and $j_{s,\omega}^{\mu}(2)$,
we will choose a local static $n^{\mu}=u^{\mu}=(1,0,0,0)$ at one
space-time point but with $\partial_{\mu}u_{\nu}\neq0$ in its vicinity,
in which we obtain the explicit form of $j_{s,\omega}^{\mu}(1)$ and
$j_{s,\omega}^{\mu}(2)$ in three spatial dimensions (3D),
\begin{eqnarray}
\mathbf{j}_{s,\omega}(1) & = & \hbar\frac{s}{2}\int\frac{d^{3}p}{(2\pi)^{3}}\frac{\mathbf{p}}{|\mathbf{p}|^{2}}(\mathbf{p}\cdot\boldsymbol{\omega})\left[f_{\mathrm{FD}}^{+}(1-f_{\mathrm{FD}}^{+})+f_{\mathrm{FD}}^{-}(1-f_{\mathrm{FD}}^{-})\right]\nonumber \\
 & \approx & \int\frac{d^{3}p}{(2\pi)^{3}}\frac{\mathbf{p}}{|\mathbf{p}|}\left[f_{\mathrm{FD}}\left(\beta|\mathbf{p}|-\beta\mu_{s}-s\hbar\frac{\mathbf{p}\cdot\boldsymbol{\omega}}{2|\mathbf{p}|}\right)\right.\nonumber \\
 &  & \left.+f_{\mathrm{FD}}\left(\beta|\mathbf{p}|+\beta\mu_{s}-s\hbar\frac{\mathbf{p}\cdot\boldsymbol{\omega}}{2|\mathbf{p}|}\right)\right],\nonumber \\
\mathbf{j}_{s,\omega}(2) & = & \lim_{|\mathbf{v}|=0}\nabla\times\int\frac{d^{3}p}{(2\pi)^{3}}\left(\frac{s\mathbf{p}}{2|\mathbf{p}|^{2}}\hbar\right)\nonumber \\
 &  & \times\left[f_{\mathrm{FD}}(\gamma\beta|\mathbf{p}|-\gamma\beta\mathbf{v}\cdot\mathbf{p}-\beta\mu_{s})+f_{\mathrm{FD}}(\gamma\beta|\mathbf{p}|-\gamma\beta\mathbf{v}\cdot\mathbf{p}+\beta\mu_{s})\right],
\end{eqnarray}
where in obtaining $\mathbf{j}_{s,\omega}(2)$ we have taken the limit
at $|\mathbf{v}|=0$ with $u^{\mu}=(\gamma,\gamma\mathbf{v})$ and
$\gamma=1/\sqrt{1-|\mathbf{v}|^{2}}$.  We can see that $\mathbf{j}_{s,\omega}(1)$
comes from the momentum integration of the fermion's velocity $\mathbf{p}/|\mathbf{p}|$
with the Fermi-Dirac distribution function in which the fermion's
energy is modified by the spin-vorticity coupling, while $\mathbf{j}_{s,\omega}(2)$
is from the magnetization due to the magnetic moment of the chiral
fermion where the magnetic moment of the chiral fermion is given by
$\hbar s\mathbf{p}/(2|\mathbf{p}|^{2})$ \cite{Chen:2014cla,Chen:2015gta,Kharzeev:2016sut}.

\textit{Acknowledgments}. QW thanks W. Florkowski for helpful discussions.
QW is supported in part by the 973 program under Grant No. 2015CB856902
and by the National Natural Science Foundation of China (NSFC) under
Grant No. 11535012. JHG is supported in part by NSFC under Grant No.
11475104, the Natural Science Foundation of Shandong Province under
the Grant No. JQ201601 and Qilu Youth Scholar Project Funding of Shandong
University. JYP is supported by the DFG (CRC 110 \textquotedblleft{}Symmetries
and the Emergence of Structure in QCD\textquotedblright{}).

\appendix

\section{Some useful formula and detailed derivations}

\label{sec:app-1}First we list useful formulas for the field strength
tensor, vorticity tensor and their duals,
\begin{eqnarray}
F^{\mu\nu} & = & E^{\mu}u^{\nu}-E^{\nu}u^{\mu}+\epsilon^{\mu\nu\rho\sigma}u_{\rho}B_{\sigma},\nonumber \\
\tilde{F}^{\mu\nu} & = & B^{\mu}u^{\nu}-B^{\nu}u^{\mu}+\epsilon^{\mu\nu\rho\sigma}E_{\rho}u_{\sigma},\nonumber \\
\tilde{F}^{\mu\nu} & = & \frac{1}{2}\epsilon^{\mu\nu\rho\lambda}F_{\rho\lambda},\nonumber \\
F^{\mu\nu} & = & -\frac{1}{2}\epsilon^{\mu\nu\rho\lambda}\tilde{F}_{\rho\lambda},\nonumber \\
\Omega^{\mu\nu} & = & \varepsilon^{\mu}u^{\nu}-\varepsilon^{\nu}u^{\mu}+\epsilon^{\mu\nu\rho\sigma}u_{\rho}\omega_{\sigma},\nonumber \\
\tilde{\Omega}^{\mu\nu} & = & \omega^{\mu}u^{\nu}-\omega^{\nu}u^{\mu}+\epsilon^{\mu\nu\rho\sigma}\varepsilon_{\rho}u_{\sigma},\nonumber \\
\Omega^{\mu\nu} & = & \frac{1}{2}(\partial^{\mu}\beta^{\nu}-\partial^{\nu}\beta^{\mu})=-\frac{1}{2}\epsilon^{\mu\nu\rho\sigma}\tilde{\Omega}_{\rho\sigma},\nonumber \\
\tilde{\Omega}^{\mu\nu} & = & \frac{1}{2}\epsilon^{\mu\nu\rho\sigma}\Omega_{\rho\sigma},\label{eq:formula}
\end{eqnarray}
where $\epsilon^{\mu\nu\sigma\beta}$ and $\epsilon_{\mu\nu\sigma\beta}$
are anti-symmetric tensors with $\epsilon^{\mu\nu\sigma\beta}=1(-1)$
and $\epsilon_{\mu\nu\sigma\beta}=-1(1)$ for even (odd) permutations
of indices 0123, so we have $\epsilon^{0123}=-\epsilon_{0123}=1$.
Instead of $\Omega_{\nu\sigma}$, $\tilde{\Omega}^{\xi\eta}$, $F_{\mu\nu}$
and $\tilde{F}^{\rho\lambda}$, usually we also use the thermal vorticity
vector $\omega^{\rho}=\frac{1}{2}\epsilon^{\rho\sigma\alpha\gamma}u_{\sigma}\partial_{\alpha}\beta_{\gamma}=\tilde{\Omega}^{\rho\sigma}u_{\sigma}$,
the electric field $E^{\mu}=F^{\mu\nu}u_{\nu}$, and the magnetic
field $B^{\mu}=\frac{1}{2}\epsilon^{\mu\nu\lambda\rho}u_{\nu}F_{\lambda\rho}$.

The contraction formula of two anti-symmetric tensors are useful
\begin{eqnarray}
\epsilon^{\gamma\mu\nu\alpha}\epsilon_{\gamma\mu\rho\xi} & = & -2(\delta_{\rho}^{\nu}\delta_{\xi}^{\alpha}-\delta_{\xi}^{\nu}\delta_{\rho}^{\alpha}),\nonumber \\
\epsilon^{\gamma\mu\nu\alpha}\epsilon_{\gamma\sigma\rho\xi} & = & -\delta_{\sigma\rho\xi}^{\mu\nu\alpha}=-\left|\begin{array}{ccc}
\delta_{\sigma}^{\mu} & \delta_{\rho}^{\mu} & \delta_{\xi}^{\mu}\\
\delta_{\sigma}^{\nu} & \delta_{\rho}^{\nu} & \delta_{\xi}^{\nu}\\
\delta_{\sigma}^{\alpha} & \delta_{\rho}^{\alpha} & \delta_{\xi}^{\alpha}
\end{array}\right|.\label{eq:epsilon}
\end{eqnarray}
With the above formula we can evaluate $\delta_{\sigma\rho\xi}^{\mu\nu\alpha}\tilde{\Omega}^{\rho\xi}p_{\alpha}p^{\sigma}$
in Eq. (\ref{eq:sol-3}),
\begin{eqnarray}
\delta_{\sigma\rho\xi}^{\mu\nu\alpha}\tilde{\Omega}^{\rho\xi}p_{\alpha}p^{\sigma} & = & \tilde{\Omega}^{\rho\xi}p_{\alpha}p^{\sigma}(\delta_{\sigma}^{\mu}\delta_{\rho}^{\nu}\delta_{\xi}^{\alpha}+\delta_{\rho}^{\mu}\delta_{\xi}^{\nu}\delta_{\sigma}^{\alpha}+\delta_{\xi}^{\mu}\delta_{\sigma}^{\nu}\delta_{\rho}^{\alpha}\nonumber \\
 &  & -\delta_{\xi}^{\mu}\delta_{\rho}^{\nu}\delta_{\sigma}^{\alpha}-\delta_{\sigma}^{\mu}\delta_{\xi}^{\nu}\delta_{\rho}^{\alpha}-\delta_{\rho}^{\mu}\delta_{\sigma}^{\nu}\delta_{\xi}^{\alpha})\nonumber \\
 & = & 2\tilde{\Omega}^{\nu\rho}p_{\rho}p^{\mu}+2\tilde{\Omega}^{\mu\nu}p^{2}-2\tilde{\Omega}^{\mu\rho}p_{\rho}p^{\nu}.\label{eq:del-o-pp}
\end{eqnarray}

We try to prove the identities in (\ref{eq:int-condition}). In order
to prove the first identity, we can start from the first term on the
right-hand side by rewriting $\Omega^{\mu\alpha}$ and $\tilde{F}_{\mu\rho}$
in linear combinations of $\tilde{\Omega}_{\lambda\gamma}$ and $F^{\eta\delta}$
respectively,
\begin{eqnarray}
\Omega^{\mu\alpha}\tilde{F}_{\mu\rho}p_{\alpha}p^{\rho} & = & -\frac{1}{4}\epsilon^{\mu\alpha\lambda\gamma}\epsilon_{\mu\rho\eta\delta}\tilde{\Omega}_{\lambda\gamma}F^{\eta\delta}p_{\alpha}p^{\rho}\nonumber \\
 & = & \frac{1}{4}\delta_{\rho\eta\delta}^{\alpha\lambda\gamma}\tilde{\Omega}_{\lambda\gamma}F^{\eta\delta}p_{\alpha}p^{\rho}\nonumber \\
 & = & \frac{1}{2}\tilde{\Omega}^{\mu\nu}F_{\mu\nu}p^{2}-\tilde{\Omega}_{\mu\rho}F^{\mu\alpha}p_{\alpha}p^{\rho},\label{eq:first-id}
\end{eqnarray}
where we have used Eq. (\ref{eq:epsilon}). For the second identity,
in the same way we can rewrite $\tilde{F}^{\mu\alpha}$ and $F_{\mu\sigma}$
on the left-hand side in linear combinations of $\tilde{F}_{\lambda\gamma}$
and $F^{\eta\delta}$ respectively,
\begin{eqnarray}
\tilde{F}^{\mu\alpha}F_{\mu\sigma}p_{\alpha}p^{\sigma} & = & F^{\mu\alpha}\tilde{F}_{\mu\rho}p_{\alpha}p^{\rho}\nonumber \\
 & = & -\frac{1}{4}\epsilon^{\mu\alpha\lambda\gamma}\epsilon_{\mu\rho\eta\delta}\tilde{F}_{\lambda\gamma}F^{\eta\delta}p_{\alpha}p^{\rho}\nonumber \\
 & = & \frac{1}{4}\delta_{\rho\eta\delta}^{\alpha\lambda\gamma}\tilde{F}_{\lambda\gamma}F^{\eta\delta}p_{\alpha}p^{\rho}\nonumber \\
 & = & \frac{1}{2}\tilde{F}^{\mu\nu}F_{\mu\nu}p^{2}-\tilde{F}_{\mu\rho}F^{\mu\alpha}p_{\alpha}p^{\rho}.\label{eq:second-id}
\end{eqnarray}
So we obtain the second identity of (\ref{eq:int-condition}). In
the same procedure we can prove the third identity of (\ref{eq:int-condition}).

We now verify $\nabla_{\mu}\mathscr{J}_{(1)}^{\mu}=0$ with $\mathscr{J}_{(1)}^{\mu}$
being given by Eq. (\ref{eq:1st-solution}). There are two parts in
$\mathscr{J}_{(1)}^{\mu}$: the vorticity part and electromagnetic
field part. We evaluate the vorticity part as
\begin{eqnarray}
\nabla_{\mu}\mathscr{J}_{(1)}^{\mu}(\Omega) & = & -\hbar\frac{s}{2}\tilde{\Omega}^{\mu\alpha}(\nabla_{\mu}p_{\alpha})f^{\prime}\delta(p^{2})-\hbar\frac{s}{2}\tilde{\Omega}^{\mu\alpha}p_{\alpha}(\nabla_{\mu}f^{\prime})\delta(p^{2})\nonumber \\
 &  & -\hbar\frac{s}{2}\tilde{\Omega}^{\mu\alpha}p_{\alpha}f^{\prime}\nabla_{\mu}\delta(p^{2})\nonumber \\
 & = & \hbar\frac{1}{2}sQ\left[\tilde{\Omega}^{\mu\alpha}F_{\mu\alpha}p^{2}-2\tilde{\Omega}^{\mu\alpha}F_{\mu\rho}p_{\alpha}p^{\rho}\right]f^{\prime}\frac{1}{p^{2}}\delta(p^{2})\nonumber \\
 &  & -\hbar\frac{s}{8}\tilde{\Omega}^{\rho\lambda}\Omega_{\rho\lambda}p^{2}f^{\prime\prime}\delta(p^{2})\nonumber \\
 & = & \hbar sQ\Omega^{\mu\alpha}\tilde{F}_{\mu\rho}p_{\alpha}p^{\rho}f^{\prime}\frac{1}{p^{2}}\delta(p^{2}),\label{eq:nabla-j-om}
\end{eqnarray}
where we have used the fact that $\tilde{\Omega}^{\mu\alpha}$ is
a constant due to the Killing condition in (\ref{eq:killing-condition}),
$\nabla_{\mu}f=f^{\prime}\Omega_{\mu\sigma}p^{\sigma}$ in Eq. (\ref{eq:nabla-f}),
and the integrability condition (\ref{eq:int-condition}). Then we
look at the electromagnetic field part
\begin{eqnarray}
\nabla_{\mu}\mathscr{J}_{(1)}^{\mu}(\mathrm{EM}) & = & \hbar sQ\tilde{F}^{\mu\lambda}(\nabla_{\mu}p_{\lambda})f\delta^{\prime}(p^{2})+\hbar sQ\tilde{F}^{\mu\lambda}p_{\lambda}(\nabla_{\mu}f)\delta^{\prime}(p^{2})\nonumber \\
 &  & +\hbar sQ\tilde{F}^{\mu\lambda}p_{\lambda}f\nabla_{\mu}\delta^{\prime}(p^{2})\nonumber \\
 & = & -\hbar sQ^{2}\tilde{F}^{\mu\lambda}F_{\mu\lambda}f\delta^{\prime}(p^{2})+\hbar sQ\tilde{F}^{\mu\lambda}\Omega_{\mu\sigma}p_{\lambda}p^{\sigma}f^{\prime}\delta^{\prime}(p^{2})\nonumber \\
 &  & +4\hbar sQ^{2}\tilde{F}^{\mu\lambda}F_{\mu\rho}p_{\lambda}p^{\rho}\frac{1}{p^{2}}f\delta^{\prime}(p^{2})\nonumber \\
 & = & \hbar sQ\tilde{F}^{\mu\lambda}\Omega_{\mu\sigma}p_{\lambda}p^{\sigma}f^{\prime}\delta^{\prime}(p^{2}),\label{eq:nabla-j-em}
\end{eqnarray}
where we have assumed $\tilde{F}^{\mu\lambda}$ is a constant and
used $\nabla_{\mu}f=f^{\prime}\Omega_{\mu\sigma}p^{\sigma}$, $\delta^{\prime}(x)=-\delta(x)/x$,
$\delta^{\prime\prime}(x)=-2\delta^{\prime}(x)/x$, $\nabla_{\mu}f=f^{\prime}\Omega_{\mu\sigma}p^{\sigma}$
in Eq. (\ref{eq:nabla-f}), and the integrability condition (\ref{eq:int-condition}).
From Eqs. (\ref{eq:nabla-j-om},\ref{eq:nabla-j-em}) we obtain $\nabla_{\mu}\mathscr{J}_{(1)}^{\mu}=0$.

\section{Derivation of Eq. (\ref{Jmu-1})}

\label{sec:app-2}For the time-like vector $n^{\mu}(x)$ with space-time
dependence Eqs. (35-38) and Eqs. (39-42) in Ref. \cite{Gao:2018wmr}
are modified by including terms with space-time derivatives of $n^{\mu}(x)$.
Note that we used $u$ in Ref. \cite{Gao:2018wmr} for the constant
reference vector while we use $n^{\mu}(x)$ for the non-constant reference
vector in the current paper. The set of equations at the zero-th order
read {[}corresponding to Eqs. (35-38) in Ref. \cite{Gao:2018wmr}{]}
\begin{eqnarray}
\bar{\nabla}\cdot\bar{\mathscr{J}}^{(0)}+(n\cdot\nabla)(n\cdot\mathscr{J}^{(0)})\nonumber \\
-\mathscr{J}_{\mu}^{(0)}(n\cdot\nabla)n^{\mu}+(n\cdot\mathscr{J}^{(0)})\nabla^{\mu}n_{\mu} & = & 0,\label{J-eq-1-0}\\
\bar{p}_{\mu}\bar{\mathscr{J}}_{\nu}^{(0)}-\bar{p}_{\nu}\bar{\mathscr{J}}_{\mu}^{(0)} & = & 0,\label{J-c2-1b-0}\\
(n\cdot p)(n\cdot\mathscr{J}^{(0)})+\bar{p}\cdot\bar{\mathscr{J}}^{(0)} & = & 0,\label{J-c1-1-0}\\
\bar{p}_{\mu}(n\cdot\mathscr{J}^{(0)})-(n\cdot p)\bar{\mathscr{J}}_{\mu}^{(0)} & = & 0,\label{J-c2-1a-0}
\end{eqnarray}
and the set of equations at the first order read {[}corresponding
to Eqs. (39-42) in Ref. \cite{Gao:2018wmr}{]}
\begin{eqnarray}
n\cdot\nabla\left(n\cdot\mathscr{J}^{(1)}\right)+\bar{\nabla}\cdot\bar{\mathscr{J}}^{(1)}\nonumber \\
-\mathscr{J}_{\mu}^{(1)}(n\cdot\nabla)n^{\mu}+(n\cdot\mathscr{J}^{(1)})\nabla^{\mu}n_{\mu} & = & 0,\label{J-eq-1-1}\\
2s\left(\bar{p}_{\mu}\bar{\mathscr{J}}_{\nu}^{(1)}-\bar{p}_{\nu}\bar{\mathscr{J}}_{\mu}^{(1)}\right) & = & -\epsilon_{\mu\nu\rho\sigma}n^{\rho}\left[(n\cdot\nabla)\bar{\mathscr{J}}^{(0)\sigma}-\bar{\nabla}^{\sigma}(n\cdot\mathscr{J}^{(0)})\right]\nonumber \\
 &  & -\epsilon_{\mu\nu\rho\sigma}u^{\rho}\mathscr{J}_{\alpha}^{(0)}\bar{\nabla}^{\sigma}u^{\alpha}-\epsilon_{\mu\nu\rho\sigma}n^{\rho}(n\cdot\mathscr{J}^{(0)})(n\cdot\nabla)n^{\sigma},\label{J-c2-1b-1}\\
(n\cdot p)(n\cdot\mathscr{J}^{(1)})+\bar{p}\cdot\bar{\mathscr{J}}^{(1)} & = & 0,\label{J-c1-1-1}\\
2s\left[\bar{p}_{\mu}(n\cdot\mathscr{J}^{(1)})-(n\cdot p)\bar{\mathscr{J}}_{\mu}^{(1)}\right] & = & -\epsilon_{\mu\nu\rho\sigma}n^{\nu}\bar{\nabla}^{\rho}\bar{\mathscr{J}}_{(0)}^{\sigma}-\epsilon_{\mu\nu\rho\sigma}(n\cdot\mathscr{J}^{(0)})n^{\nu}\bar{\nabla}^{\rho}n^{\sigma}.\label{J-c2-1a-1}
\end{eqnarray}
From Eq. (\ref{J-c2-1a-0}) and Eq. (\ref{J-c2-1a-1}), we obtain
\begin{eqnarray}
\bar{\mathscr{J}}_{\mu}^{(0)} & = & \bar{p}_{\mu}\frac{n\cdot\mathscr{J}^{(0)}}{n\cdot p},\nonumber \\
\bar{\mathscr{J}}_{\mu}^{(1)} & = & \bar{p}_{\mu}\frac{n\cdot\mathscr{J}^{(1)}}{n\cdot p}-\frac{s}{2(n\cdot p)}\epsilon_{\mu\nu\rho\sigma}n^{\nu}\nabla^{\sigma}\bar{\mathscr{J}}_{(0)}^{\rho}-\frac{s(n\cdot\mathscr{J}^{(0)})}{2(n\cdot p)}\epsilon_{\mu\nu\rho\sigma}n^{\nu}\nabla^{\rho}n^{\sigma}\nonumber \\
 & = & \bar{p}_{\mu}\frac{n\cdot\mathscr{J}^{(1)}}{n\cdot p}-\frac{s}{2(n\cdot p)}\epsilon_{\mu\nu\rho\sigma}n^{\nu}\nabla^{\sigma}\mathscr{J}_{(0)}^{\rho}.\label{eq:j-bar-1}
\end{eqnarray}
We see in $\bar{\mathscr{J}}_{\mu}^{(1)}$ that there is a term proportional
to $\epsilon_{\mu\nu\rho\sigma}n^{\nu}\nabla^{\rho}n^{\sigma}$ which
is vanishing for a constant $n^{\mu}$, in the case the term $\epsilon_{\mu\nu\rho\sigma}n^{\nu}\nabla^{\sigma}\bar{\mathscr{J}}_{(0)}^{\rho}$
would be $\epsilon_{\mu\nu\rho\sigma}n^{\nu}\nabla^{\sigma}\mathscr{J}_{(0)}^{\rho}$
and we immediately obtain second equality in $\bar{\mathscr{J}}_{\mu}^{(1)}$.
But for $n^{\mu}(x)$ as a function of space-time, one has to evaluate
both terms, $\epsilon_{\mu\nu\rho\sigma}n^{\nu}\nabla^{\sigma}\bar{\mathscr{J}}_{(0)}^{\rho}$
and $\epsilon_{\mu\nu\rho\sigma}n^{\nu}\nabla^{\rho}n^{\sigma}$,
to reach the second equality. We combine the components parallel and
orthorgonal to $n^{\mu}$ to obtain
\begin{eqnarray}
\mathscr{J}_{\mu}^{(0)} & = & p_{\mu}\frac{n\cdot\mathscr{J}^{(0)}}{n\cdot p},\nonumber \\
\mathscr{J}_{\mu}^{(1)} & = & p_{\mu}\frac{n\cdot\mathscr{J}^{(1)}}{n\cdot p}-\frac{s}{2(n\cdot p)}\epsilon_{\mu\nu\rho\sigma}n^{\nu}\nabla^{\sigma}\mathscr{J}_{(0)}^{\rho}.\label{eq:j0j1}
\end{eqnarray}
Using the mass-shell conditions (\ref{J-c1-1-0}) and (\ref{J-c1-1-1})
we obtain the general solution,
\begin{eqnarray}
\frac{u\cdot\mathscr{J}^{(0)}}{u\cdot p} & = & f^{(0)}\delta\left(p^{2}\right),\\
\frac{u\cdot\mathscr{J}^{(1)}}{u\cdot p} & = & f^{(1)}\delta\left(p^{2}\right)-sQ\frac{B\cdot p}{u\cdot p}f^{(0)}\delta^{\prime}\left(p^{2}\right).\label{eq:j1-1}
\end{eqnarray}
When neglecting the electromagnetic field, we obtain Eq. (\ref{Jmu-1})
from Eqs. (\ref{eq:j0j1}-\ref{eq:j1-1}).

\bibliographystyle{apsrev}
\bibliography{ref-1}

\begin{thebibliography}{67}
\expandafter\ifx\csname natexlab\endcsname\relax\def\natexlab#1{#1}\fi
\expandafter\ifx\csname bibnamefont\endcsname\relax
  \def\bibnamefont#1{#1}\fi
\expandafter\ifx\csname bibfnamefont\endcsname\relax
  \def\bibfnamefont#1{#1}\fi
\expandafter\ifx\csname citenamefont\endcsname\relax
  \def\citenamefont#1{#1}\fi
\expandafter\ifx\csname url\endcsname\relax
  \def\url#1{\texttt{#1}}\fi
\expandafter\ifx\csname urlprefix\endcsname\relax\def\urlprefix{URL }\fi
\providecommand{\bibinfo}[2]{#2}
\providecommand{\eprint}[2][]{\url{#2}}

\bibitem[{\citenamefont{Gao et~al.}(2012)\citenamefont{Gao, Liang, Pu, Wang,
  and Wang}}]{Gao:2012ix}
\bibinfo{author}{\bibfnamefont{J.-H.} \bibnamefont{Gao}},
  \bibinfo{author}{\bibfnamefont{Z.-T.} \bibnamefont{Liang}},
  \bibinfo{author}{\bibfnamefont{S.}~\bibnamefont{Pu}},
  \bibinfo{author}{\bibfnamefont{Q.}~\bibnamefont{Wang}}, \bibnamefont{and}
  \bibinfo{author}{\bibfnamefont{X.-N.} \bibnamefont{Wang}},
  \bibinfo{journal}{Phys.Rev.Lett.} \textbf{\bibinfo{volume}{109}},
  \bibinfo{pages}{232301} (\bibinfo{year}{2012}), \eprint{1203.0725}.

\bibitem[{\citenamefont{Gao et~al.}(2018)\citenamefont{Gao, Liang, Wang, and
  Wang}}]{Gao:2018wmr}
\bibinfo{author}{\bibfnamefont{J.-H.} \bibnamefont{Gao}},
  \bibinfo{author}{\bibfnamefont{Z.-T.} \bibnamefont{Liang}},
  \bibinfo{author}{\bibfnamefont{Q.}~\bibnamefont{Wang}}, \bibnamefont{and}
  \bibinfo{author}{\bibfnamefont{X.-N.} \bibnamefont{Wang}},
  \bibinfo{journal}{Phys. Rev.} \textbf{\bibinfo{volume}{D98}},
  \bibinfo{pages}{036019} (\bibinfo{year}{2018}), \eprint{1802.06216}.

\bibitem[{\citenamefont{Einstein and de~Haas}(1915)}]{dehaas:1915}
\bibinfo{author}{\bibfnamefont{A.}~\bibnamefont{Einstein}} \bibnamefont{and}
  \bibinfo{author}{\bibfnamefont{W.}~\bibnamefont{de~Haas}},
  \bibinfo{journal}{Deutsche Physikalische Gesellschaft, Verhandlungen}
  \textbf{\bibinfo{volume}{17}}, \bibinfo{pages}{152} (\bibinfo{year}{1915}).

\bibitem[{\citenamefont{Barnett}(1935)}]{Barnett:1935}
\bibinfo{author}{\bibfnamefont{S.}~\bibnamefont{Barnett}},
  \bibinfo{journal}{Rev. Mod. Rev.} \textbf{\bibinfo{volume}{7}},
  \bibinfo{pages}{129} (\bibinfo{year}{1935}).

\bibitem[{\citenamefont{Liang and Wang}(2005{\natexlab{a}})}]{Liang:2004ph}
\bibinfo{author}{\bibfnamefont{Z.-T.} \bibnamefont{Liang}} \bibnamefont{and}
  \bibinfo{author}{\bibfnamefont{X.-N.} \bibnamefont{Wang}},
  \bibinfo{journal}{Phys. Rev. Lett.} \textbf{\bibinfo{volume}{94}},
  \bibinfo{pages}{102301} (\bibinfo{year}{2005}{\natexlab{a}}),
  \bibinfo{note}{[Erratum: Phys. Rev. Lett.96,039901(2006)]},
  \eprint{nucl-th/0410079}.

\bibitem[{\citenamefont{Liang and Wang}(2005{\natexlab{b}})}]{Liang:2004xn}
\bibinfo{author}{\bibfnamefont{Z.-T.} \bibnamefont{Liang}} \bibnamefont{and}
  \bibinfo{author}{\bibfnamefont{X.-N.} \bibnamefont{Wang}},
  \bibinfo{journal}{Phys. Lett.} \textbf{\bibinfo{volume}{B629}},
  \bibinfo{pages}{20} (\bibinfo{year}{2005}{\natexlab{b}}),
  \eprint{nucl-th/0411101}.

\bibitem[{\citenamefont{Voloshin}(2004)}]{Voloshin:2004ha}
\bibinfo{author}{\bibfnamefont{S.~A.} \bibnamefont{Voloshin}}
  (\bibinfo{year}{2004}), \eprint{nucl-th/0410089}.

\bibitem[{\citenamefont{Betz et~al.}(2007)\citenamefont{Betz, Gyulassy, and
  Torrieri}}]{Betz:2007kg}
\bibinfo{author}{\bibfnamefont{B.}~\bibnamefont{Betz}},
  \bibinfo{author}{\bibfnamefont{M.}~\bibnamefont{Gyulassy}}, \bibnamefont{and}
  \bibinfo{author}{\bibfnamefont{G.}~\bibnamefont{Torrieri}},
  \bibinfo{journal}{Phys. Rev.} \textbf{\bibinfo{volume}{C76}},
  \bibinfo{pages}{044901} (\bibinfo{year}{2007}), \eprint{0708.0035}.

\bibitem[{\citenamefont{Becattini et~al.}(2008)\citenamefont{Becattini,
  Piccinini, and Rizzo}}]{Becattini:2007sr}
\bibinfo{author}{\bibfnamefont{F.}~\bibnamefont{Becattini}},
  \bibinfo{author}{\bibfnamefont{F.}~\bibnamefont{Piccinini}},
  \bibnamefont{and} \bibinfo{author}{\bibfnamefont{J.}~\bibnamefont{Rizzo}},
  \bibinfo{journal}{Phys. Rev.} \textbf{\bibinfo{volume}{C77}},
  \bibinfo{pages}{024906} (\bibinfo{year}{2008}), \eprint{0711.1253}.

\bibitem[{\citenamefont{Gao et~al.}(2008)\citenamefont{Gao, Chen, Deng, Liang,
  Wang, and Wang}}]{Gao:2007bc}
\bibinfo{author}{\bibfnamefont{J.-H.} \bibnamefont{Gao}},
  \bibinfo{author}{\bibfnamefont{S.-W.} \bibnamefont{Chen}},
  \bibinfo{author}{\bibfnamefont{W.-t.} \bibnamefont{Deng}},
  \bibinfo{author}{\bibfnamefont{Z.-T.} \bibnamefont{Liang}},
  \bibinfo{author}{\bibfnamefont{Q.}~\bibnamefont{Wang}}, \bibnamefont{and}
  \bibinfo{author}{\bibfnamefont{X.-N.} \bibnamefont{Wang}},
  \bibinfo{journal}{Phys. Rev.} \textbf{\bibinfo{volume}{C77}},
  \bibinfo{pages}{044902} (\bibinfo{year}{2008}), \eprint{0710.2943}.

\bibitem[{\citenamefont{Adamczyk et~al.}(2017)}]{STAR:2017ckg}
\bibinfo{author}{\bibfnamefont{L.}~\bibnamefont{Adamczyk}} \bibnamefont{et~al.}
  (\bibinfo{collaboration}{STAR}) (\bibinfo{year}{2017}), \eprint{1701.06657}.

\bibitem[{\citenamefont{Becattini et~al.}(2013)\citenamefont{Becattini,
  Chandra, Del~Zanna, and Grossi}}]{Becattini:2013fla}
\bibinfo{author}{\bibfnamefont{F.}~\bibnamefont{Becattini}},
  \bibinfo{author}{\bibfnamefont{V.}~\bibnamefont{Chandra}},
  \bibinfo{author}{\bibfnamefont{L.}~\bibnamefont{Del~Zanna}},
  \bibnamefont{and} \bibinfo{author}{\bibfnamefont{E.}~\bibnamefont{Grossi}},
  \bibinfo{journal}{Annals Phys.} \textbf{\bibinfo{volume}{338}},
  \bibinfo{pages}{32} (\bibinfo{year}{2013}), \eprint{1303.3431}.

\bibitem[{\citenamefont{Becattini and Grossi}(2015)}]{Becattini:2015nva}
\bibinfo{author}{\bibfnamefont{F.}~\bibnamefont{Becattini}} \bibnamefont{and}
  \bibinfo{author}{\bibfnamefont{E.}~\bibnamefont{Grossi}},
  \bibinfo{journal}{Phys. Rev.} \textbf{\bibinfo{volume}{D92}},
  \bibinfo{pages}{045037} (\bibinfo{year}{2015}), \eprint{1505.07760}.

\bibitem[{\citenamefont{Becattini et~al.}(2016)\citenamefont{Becattini,
  Karpenko, Lisa, Upsal, and Voloshin}}]{Becattini:2016gvu}
\bibinfo{author}{\bibfnamefont{F.}~\bibnamefont{Becattini}},
  \bibinfo{author}{\bibfnamefont{I.}~\bibnamefont{Karpenko}},
  \bibinfo{author}{\bibfnamefont{M.}~\bibnamefont{Lisa}},
  \bibinfo{author}{\bibfnamefont{I.}~\bibnamefont{Upsal}}, \bibnamefont{and}
  \bibinfo{author}{\bibfnamefont{S.}~\bibnamefont{Voloshin}}
  (\bibinfo{year}{2016}), \eprint{1610.02506}.

\bibitem[{\citenamefont{Fang et~al.}(2016)\citenamefont{Fang, Pang, Wang, and
  Wang}}]{Fang:2016vpj}
\bibinfo{author}{\bibfnamefont{R.-h.} \bibnamefont{Fang}},
  \bibinfo{author}{\bibfnamefont{L.-g.} \bibnamefont{Pang}},
  \bibinfo{author}{\bibfnamefont{Q.}~\bibnamefont{Wang}}, \bibnamefont{and}
  \bibinfo{author}{\bibfnamefont{X.-n.} \bibnamefont{Wang}},
  \bibinfo{journal}{Phys. Rev.} \textbf{\bibinfo{volume}{C94}},
  \bibinfo{pages}{024904} (\bibinfo{year}{2016}), \eprint{1604.04036}.

\bibitem[{\citenamefont{Karpenko and Becattini}(2016)}]{Karpenko:2016jyx}
\bibinfo{author}{\bibfnamefont{I.}~\bibnamefont{Karpenko}} \bibnamefont{and}
  \bibinfo{author}{\bibfnamefont{F.}~\bibnamefont{Becattini}}
  (\bibinfo{year}{2016}), \eprint{1610.04717}.

\bibitem[{\citenamefont{Xie et~al.}(2017)\citenamefont{Xie, Wang, and
  Csernai}}]{Xie:2017upb}
\bibinfo{author}{\bibfnamefont{Y.}~\bibnamefont{Xie}},
  \bibinfo{author}{\bibfnamefont{D.}~\bibnamefont{Wang}}, \bibnamefont{and}
  \bibinfo{author}{\bibfnamefont{L.~P.} \bibnamefont{Csernai}},
  \bibinfo{journal}{Phys. Rev.} \textbf{\bibinfo{volume}{C95}},
  \bibinfo{pages}{031901} (\bibinfo{year}{2017}), \eprint{1703.03770}.

\bibitem[{\citenamefont{Li et~al.}(2017)\citenamefont{Li, Pang, Wang, and
  Xia}}]{Li:2017slc}
\bibinfo{author}{\bibfnamefont{H.}~\bibnamefont{Li}},
  \bibinfo{author}{\bibfnamefont{L.-G.} \bibnamefont{Pang}},
  \bibinfo{author}{\bibfnamefont{Q.}~\bibnamefont{Wang}}, \bibnamefont{and}
  \bibinfo{author}{\bibfnamefont{X.-L.} \bibnamefont{Xia}},
  \bibinfo{journal}{Phys. Rev.} \textbf{\bibinfo{volume}{C96}},
  \bibinfo{pages}{054908} (\bibinfo{year}{2017}), \eprint{1704.01507}.

\bibitem[{\citenamefont{Sun and Ko}(2017)}]{Sun:2017xhx}
\bibinfo{author}{\bibfnamefont{Y.}~\bibnamefont{Sun}} \bibnamefont{and}
  \bibinfo{author}{\bibfnamefont{C.~M.} \bibnamefont{Ko}},
  \bibinfo{journal}{Phys. Rev.} \textbf{\bibinfo{volume}{C96}},
  \bibinfo{pages}{024906} (\bibinfo{year}{2017}), \eprint{1706.09467}.

\bibitem[{\citenamefont{Shi et~al.}(2017)\citenamefont{Shi, Li, and
  Liao}}]{Shi:2017wpk}
\bibinfo{author}{\bibfnamefont{S.}~\bibnamefont{Shi}},
  \bibinfo{author}{\bibfnamefont{K.}~\bibnamefont{Li}}, \bibnamefont{and}
  \bibinfo{author}{\bibfnamefont{J.}~\bibnamefont{Liao}}
  (\bibinfo{year}{2017}), \eprint{1712.00878}.

\bibitem[{\citenamefont{Baznat et~al.}(2013)\citenamefont{Baznat, Gudima,
  Sorin, and Teryaev}}]{Baznat:2013zx}
\bibinfo{author}{\bibfnamefont{M.}~\bibnamefont{Baznat}},
  \bibinfo{author}{\bibfnamefont{K.}~\bibnamefont{Gudima}},
  \bibinfo{author}{\bibfnamefont{A.}~\bibnamefont{Sorin}}, \bibnamefont{and}
  \bibinfo{author}{\bibfnamefont{O.}~\bibnamefont{Teryaev}},
  \bibinfo{journal}{Phys. Rev.} \textbf{\bibinfo{volume}{C88}},
  \bibinfo{pages}{061901} (\bibinfo{year}{2013}), \eprint{1301.7003}.

\bibitem[{\citenamefont{Csernai et~al.}(2013)\citenamefont{Csernai, Magas, and
  Wang}}]{Csernai:2013bqa}
\bibinfo{author}{\bibfnamefont{L.~P.} \bibnamefont{Csernai}},
  \bibinfo{author}{\bibfnamefont{V.~K.} \bibnamefont{Magas}}, \bibnamefont{and}
  \bibinfo{author}{\bibfnamefont{D.~J.} \bibnamefont{Wang}},
  \bibinfo{journal}{Phys. Rev.} \textbf{\bibinfo{volume}{C87}},
  \bibinfo{pages}{034906} (\bibinfo{year}{2013}), \eprint{1302.5310}.

\bibitem[{\citenamefont{Csernai et~al.}(2014)\citenamefont{Csernai, Wang,
  Bleicher, and Stoecker}}]{Csernai:2014ywa}
\bibinfo{author}{\bibfnamefont{L.~P.} \bibnamefont{Csernai}},
  \bibinfo{author}{\bibfnamefont{D.~J.} \bibnamefont{Wang}},
  \bibinfo{author}{\bibfnamefont{M.}~\bibnamefont{Bleicher}}, \bibnamefont{and}
  \bibinfo{author}{\bibfnamefont{H.}~\bibnamefont{Stoecker}},
  \bibinfo{journal}{Phys. Rev.} \textbf{\bibinfo{volume}{C90}},
  \bibinfo{pages}{021904} (\bibinfo{year}{2014}).

\bibitem[{\citenamefont{Teryaev and Usubov}(2015)}]{Teryaev:2015gxa}
\bibinfo{author}{\bibfnamefont{O.}~\bibnamefont{Teryaev}} \bibnamefont{and}
  \bibinfo{author}{\bibfnamefont{R.}~\bibnamefont{Usubov}},
  \bibinfo{journal}{Phys. Rev.} \textbf{\bibinfo{volume}{C92}},
  \bibinfo{pages}{014906} (\bibinfo{year}{2015}).

\bibitem[{\citenamefont{Jiang et~al.}(2016)\citenamefont{Jiang, Lin, and
  Liao}}]{Jiang:2016woz}
\bibinfo{author}{\bibfnamefont{Y.}~\bibnamefont{Jiang}},
  \bibinfo{author}{\bibfnamefont{Z.-W.} \bibnamefont{Lin}}, \bibnamefont{and}
  \bibinfo{author}{\bibfnamefont{J.}~\bibnamefont{Liao}},
  \bibinfo{journal}{Phys. Rev.} \textbf{\bibinfo{volume}{C94}},
  \bibinfo{pages}{044910} (\bibinfo{year}{2016}), \bibinfo{note}{[Erratum:
  Phys. Rev.C95,no.4,049904(2017)]}, \eprint{1602.06580}.

\bibitem[{\citenamefont{Deng and Huang}(2016)}]{Deng:2016gyh}
\bibinfo{author}{\bibfnamefont{W.-T.} \bibnamefont{Deng}} \bibnamefont{and}
  \bibinfo{author}{\bibfnamefont{X.-G.} \bibnamefont{Huang}},
  \bibinfo{journal}{Phys. Rev.} \textbf{\bibinfo{volume}{C93}},
  \bibinfo{pages}{064907} (\bibinfo{year}{2016}), \eprint{1603.06117}.

\bibitem[{\citenamefont{Ivanov and Soldatov}(2017)}]{Ivanov:2017dff}
\bibinfo{author}{\bibfnamefont{{\relax Yu}.~B.} \bibnamefont{Ivanov}}
  \bibnamefont{and} \bibinfo{author}{\bibfnamefont{A.~A.}
  \bibnamefont{Soldatov}}, \bibinfo{journal}{Phys. Rev.}
  \textbf{\bibinfo{volume}{C95}}, \bibinfo{pages}{054915}
  (\bibinfo{year}{2017}), \eprint{1701.01319}.

\bibitem[{\citenamefont{Becattini et~al.}(2015)\citenamefont{Becattini,
  Inghirami, Rolando, Beraudo, Del~Zanna, De~Pace, Nardi, Pagliara, and
  Chandra}}]{Becattini:2015ska}
\bibinfo{author}{\bibfnamefont{F.}~\bibnamefont{Becattini}},
  \bibinfo{author}{\bibfnamefont{G.}~\bibnamefont{Inghirami}},
  \bibinfo{author}{\bibfnamefont{V.}~\bibnamefont{Rolando}},
  \bibinfo{author}{\bibfnamefont{A.}~\bibnamefont{Beraudo}},
  \bibinfo{author}{\bibfnamefont{L.}~\bibnamefont{Del~Zanna}},
  \bibinfo{author}{\bibfnamefont{A.}~\bibnamefont{De~Pace}},
  \bibinfo{author}{\bibfnamefont{M.}~\bibnamefont{Nardi}},
  \bibinfo{author}{\bibfnamefont{G.}~\bibnamefont{Pagliara}}, \bibnamefont{and}
  \bibinfo{author}{\bibfnamefont{V.}~\bibnamefont{Chandra}},
  \bibinfo{journal}{Eur. Phys. J.} \textbf{\bibinfo{volume}{C75}},
  \bibinfo{pages}{406} (\bibinfo{year}{2015}), \bibinfo{note}{[Erratum: Eur.
  Phys. J.C78,no.5,354(2018)]}, \eprint{1501.04468}.

\bibitem[{\citenamefont{Becattini and Karpenko}(2018)}]{Becattini:2017gcx}
\bibinfo{author}{\bibfnamefont{F.}~\bibnamefont{Becattini}} \bibnamefont{and}
  \bibinfo{author}{\bibfnamefont{I.}~\bibnamefont{Karpenko}},
  \bibinfo{journal}{Phys. Rev. Lett.} \textbf{\bibinfo{volume}{120}},
  \bibinfo{pages}{012302} (\bibinfo{year}{2018}), \eprint{1707.07984}.

\bibitem[{\citenamefont{Karpenko and Becattini}(2017)}]{Karpenko:2017dui}
\bibinfo{author}{\bibfnamefont{I.}~\bibnamefont{Karpenko}} \bibnamefont{and}
  \bibinfo{author}{\bibfnamefont{F.}~\bibnamefont{Becattini}}
  (\bibinfo{year}{2017}), \bibinfo{note}{[EPJ Web Conf.171,17001(2018)]},
  \eprint{1710.09726}.

\bibitem[{\citenamefont{Voloshin}(2017)}]{Voloshin:2017kqp}
\bibinfo{author}{\bibfnamefont{S.~A.} \bibnamefont{Voloshin}}
  (\bibinfo{year}{2017}), \bibinfo{note}{[EPJ Web Conf.17,10700(2018)]},
  \eprint{1710.08934}.

\bibitem[{\citenamefont{Xia et~al.}(2018)\citenamefont{Xia, Li, Tang, and
  Wang}}]{Xia:2018tes}
\bibinfo{author}{\bibfnamefont{X.-L.} \bibnamefont{Xia}},
  \bibinfo{author}{\bibfnamefont{H.}~\bibnamefont{Li}},
  \bibinfo{author}{\bibfnamefont{Z.-B.} \bibnamefont{Tang}}, \bibnamefont{and}
  \bibinfo{author}{\bibfnamefont{Q.}~\bibnamefont{Wang}}
  (\bibinfo{year}{2018}), \eprint{1803.00867}.

\bibitem[{\citenamefont{Florkowski et~al.}(2017)\citenamefont{Florkowski,
  Friman, Jaiswal, Ryblewski, and Speranza}}]{Florkowski:2017dyn}
\bibinfo{author}{\bibfnamefont{W.}~\bibnamefont{Florkowski}},
  \bibinfo{author}{\bibfnamefont{B.}~\bibnamefont{Friman}},
  \bibinfo{author}{\bibfnamefont{A.}~\bibnamefont{Jaiswal}},
  \bibinfo{author}{\bibfnamefont{R.}~\bibnamefont{Ryblewski}},
  \bibnamefont{and} \bibinfo{author}{\bibfnamefont{E.}~\bibnamefont{Speranza}}
  (\bibinfo{year}{2017}), \eprint{1712.07676}.

\bibitem[{\citenamefont{Florkowski et~al.}(2018)\citenamefont{Florkowski,
  Friman, Jaiswal, and Speranza}}]{Florkowski:2017ruc}
\bibinfo{author}{\bibfnamefont{W.}~\bibnamefont{Florkowski}},
  \bibinfo{author}{\bibfnamefont{B.}~\bibnamefont{Friman}},
  \bibinfo{author}{\bibfnamefont{A.}~\bibnamefont{Jaiswal}}, \bibnamefont{and}
  \bibinfo{author}{\bibfnamefont{E.}~\bibnamefont{Speranza}},
  \bibinfo{journal}{Phys. Rev.} \textbf{\bibinfo{volume}{C97}},
  \bibinfo{pages}{041901} (\bibinfo{year}{2018}), \eprint{1705.00587}.

\bibitem[{\citenamefont{Heinz}(1983)}]{Heinz:1983nx}
\bibinfo{author}{\bibfnamefont{U.~W.} \bibnamefont{Heinz}},
  \bibinfo{journal}{Phys. Rev. Lett.} \textbf{\bibinfo{volume}{51}},
  \bibinfo{pages}{351} (\bibinfo{year}{1983}).

\bibitem[{\citenamefont{Elze et~al.}(1986)\citenamefont{Elze, Gyulassy, and
  Vasak}}]{Elze:1986qd}
\bibinfo{author}{\bibfnamefont{H.-T.} \bibnamefont{Elze}},
  \bibinfo{author}{\bibfnamefont{M.}~\bibnamefont{Gyulassy}}, \bibnamefont{and}
  \bibinfo{author}{\bibfnamefont{D.}~\bibnamefont{Vasak}},
  \bibinfo{journal}{Nucl.Phys.} \textbf{\bibinfo{volume}{B276}},
  \bibinfo{pages}{706} (\bibinfo{year}{1986}).

\bibitem[{\citenamefont{Vasak et~al.}(1987)\citenamefont{Vasak, Gyulassy, and
  Elze}}]{Vasak:1987um}
\bibinfo{author}{\bibfnamefont{D.}~\bibnamefont{Vasak}},
  \bibinfo{author}{\bibfnamefont{M.}~\bibnamefont{Gyulassy}}, \bibnamefont{and}
  \bibinfo{author}{\bibfnamefont{H.-T.} \bibnamefont{Elze}},
  \bibinfo{journal}{Annals Phys.} \textbf{\bibinfo{volume}{173}},
  \bibinfo{pages}{462} (\bibinfo{year}{1987}).

\bibitem[{\citenamefont{Zhuang and Heinz}(1996)}]{Zhuang:1995pd}
\bibinfo{author}{\bibfnamefont{P.}~\bibnamefont{Zhuang}} \bibnamefont{and}
  \bibinfo{author}{\bibfnamefont{U.~W.} \bibnamefont{Heinz}},
  \bibinfo{journal}{Annals Phys.} \textbf{\bibinfo{volume}{245}},
  \bibinfo{pages}{311} (\bibinfo{year}{1996}), \eprint{nucl-th/9502034}.

\bibitem[{\citenamefont{Florkowski et~al.}(1996)\citenamefont{Florkowski,
  Hufner, Klevansky, and Neise}}]{Florkowski:1995ei}
\bibinfo{author}{\bibfnamefont{W.}~\bibnamefont{Florkowski}},
  \bibinfo{author}{\bibfnamefont{J.}~\bibnamefont{Hufner}},
  \bibinfo{author}{\bibfnamefont{S.~P.} \bibnamefont{Klevansky}},
  \bibnamefont{and} \bibinfo{author}{\bibfnamefont{L.}~\bibnamefont{Neise}},
  \bibinfo{journal}{Annals Phys.} \textbf{\bibinfo{volume}{245}},
  \bibinfo{pages}{445} (\bibinfo{year}{1996}), \eprint{hep-ph/9505407}.

\bibitem[{\citenamefont{Blaizot and Iancu}(2002)}]{Blaizot:2001nr}
\bibinfo{author}{\bibfnamefont{J.-P.} \bibnamefont{Blaizot}} \bibnamefont{and}
  \bibinfo{author}{\bibfnamefont{E.}~\bibnamefont{Iancu}},
  \bibinfo{journal}{Phys. Rept.} \textbf{\bibinfo{volume}{359}},
  \bibinfo{pages}{355} (\bibinfo{year}{2002}), \eprint{hep-ph/0101103}.

\bibitem[{\citenamefont{Wang et~al.}(2002)\citenamefont{Wang, Redlich,
  Stoecker, and Greiner}}]{Wang:2001dm}
\bibinfo{author}{\bibfnamefont{Q.}~\bibnamefont{Wang}},
  \bibinfo{author}{\bibfnamefont{K.}~\bibnamefont{Redlich}},
  \bibinfo{author}{\bibfnamefont{H.}~\bibnamefont{Stoecker}}, \bibnamefont{and}
  \bibinfo{author}{\bibfnamefont{W.}~\bibnamefont{Greiner}},
  \bibinfo{journal}{Phys. Rev. Lett.} \textbf{\bibinfo{volume}{88}},
  \bibinfo{pages}{132303} (\bibinfo{year}{2002}), \eprint{nucl-th/0111040}.

\bibitem[{\citenamefont{Vilenkin}(1980)}]{Vilenkin:1980fu}
\bibinfo{author}{\bibfnamefont{A.}~\bibnamefont{Vilenkin}},
  \bibinfo{journal}{Phys. Rev.} \textbf{\bibinfo{volume}{D22}},
  \bibinfo{pages}{3080} (\bibinfo{year}{1980}).

\bibitem[{\citenamefont{Kharzeev et~al.}(2008)\citenamefont{Kharzeev, McLerran,
  and Warringa}}]{Kharzeev:2007jp}
\bibinfo{author}{\bibfnamefont{D.~E.} \bibnamefont{Kharzeev}},
  \bibinfo{author}{\bibfnamefont{L.~D.} \bibnamefont{McLerran}},
  \bibnamefont{and} \bibinfo{author}{\bibfnamefont{H.~J.}
  \bibnamefont{Warringa}}, \bibinfo{journal}{Nucl.Phys.}
  \textbf{\bibinfo{volume}{A803}}, \bibinfo{pages}{227} (\bibinfo{year}{2008}),
  \eprint{0711.0950}.

\bibitem[{\citenamefont{Fukushima et~al.}(2008)\citenamefont{Fukushima,
  Kharzeev, and Warringa}}]{Fukushima:2008xe}
\bibinfo{author}{\bibfnamefont{K.}~\bibnamefont{Fukushima}},
  \bibinfo{author}{\bibfnamefont{D.~E.} \bibnamefont{Kharzeev}},
  \bibnamefont{and} \bibinfo{author}{\bibfnamefont{H.~J.}
  \bibnamefont{Warringa}}, \bibinfo{journal}{Phys.Rev.}
  \textbf{\bibinfo{volume}{D78}}, \bibinfo{pages}{074033}
  (\bibinfo{year}{2008}), \eprint{0808.3382}.

\bibitem[{\citenamefont{Kharzeev
  et~al.}(2016{\natexlab{a}})\citenamefont{Kharzeev, Liao, Voloshin, and
  Wang}}]{Kharzeev:2015znc}
\bibinfo{author}{\bibfnamefont{D.~E.} \bibnamefont{Kharzeev}},
  \bibinfo{author}{\bibfnamefont{J.}~\bibnamefont{Liao}},
  \bibinfo{author}{\bibfnamefont{S.~A.} \bibnamefont{Voloshin}},
  \bibnamefont{and} \bibinfo{author}{\bibfnamefont{G.}~\bibnamefont{Wang}},
  \bibinfo{journal}{Prog. Part. Nucl. Phys.} \textbf{\bibinfo{volume}{88}},
  \bibinfo{pages}{1} (\bibinfo{year}{2016}{\natexlab{a}}), \eprint{1511.04050}.

\bibitem[{\citenamefont{Kharzeev et~al.}(2013)\citenamefont{Kharzeev,
  Landsteiner, Schmitt, and Yee}}]{Kharzeev:2013jha}
\bibinfo{author}{\bibfnamefont{D.}~\bibnamefont{Kharzeev}},
  \bibinfo{author}{\bibfnamefont{K.}~\bibnamefont{Landsteiner}},
  \bibinfo{author}{\bibfnamefont{A.}~\bibnamefont{Schmitt}}, \bibnamefont{and}
  \bibinfo{author}{\bibfnamefont{H.-U.} \bibnamefont{Yee}},
  \bibinfo{journal}{Lect. Notes Phys.} \textbf{\bibinfo{volume}{871}},
  \bibinfo{pages}{pp.1} (\bibinfo{year}{2013}).

\bibitem[{\citenamefont{Huang}(2016)}]{Huang:2015oca}
\bibinfo{author}{\bibfnamefont{X.-G.} \bibnamefont{Huang}},
  \bibinfo{journal}{Rept. Prog. Phys.} \textbf{\bibinfo{volume}{79}},
  \bibinfo{pages}{076302} (\bibinfo{year}{2016}), \eprint{1509.04073}.

\bibitem[{\citenamefont{Vilenkin}(1978)}]{Vilenkin:1978hb}
\bibinfo{author}{\bibfnamefont{A.}~\bibnamefont{Vilenkin}},
  \bibinfo{journal}{Phys. Lett.} \textbf{\bibinfo{volume}{B80}},
  \bibinfo{pages}{150} (\bibinfo{year}{1978}).

\bibitem[{\citenamefont{Erdmenger et~al.}(2009)\citenamefont{Erdmenger, Haack,
  Kaminski, and Yarom}}]{Erdmenger:2008rm}
\bibinfo{author}{\bibfnamefont{J.}~\bibnamefont{Erdmenger}},
  \bibinfo{author}{\bibfnamefont{M.}~\bibnamefont{Haack}},
  \bibinfo{author}{\bibfnamefont{M.}~\bibnamefont{Kaminski}}, \bibnamefont{and}
  \bibinfo{author}{\bibfnamefont{A.}~\bibnamefont{Yarom}},
  \bibinfo{journal}{JHEP} \textbf{\bibinfo{volume}{01}}, \bibinfo{pages}{055}
  (\bibinfo{year}{2009}), \eprint{0809.2488}.

\bibitem[{\citenamefont{Banerjee et~al.}(2011)\citenamefont{Banerjee,
  Bhattacharya, Bhattacharyya, Dutta, Loganayagam, and
  Surowka}}]{Banerjee:2008th}
\bibinfo{author}{\bibfnamefont{N.}~\bibnamefont{Banerjee}},
  \bibinfo{author}{\bibfnamefont{J.}~\bibnamefont{Bhattacharya}},
  \bibinfo{author}{\bibfnamefont{S.}~\bibnamefont{Bhattacharyya}},
  \bibinfo{author}{\bibfnamefont{S.}~\bibnamefont{Dutta}},
  \bibinfo{author}{\bibfnamefont{R.}~\bibnamefont{Loganayagam}},
  \bibnamefont{and} \bibinfo{author}{\bibfnamefont{P.}~\bibnamefont{Surowka}},
  \bibinfo{journal}{JHEP} \textbf{\bibinfo{volume}{01}}, \bibinfo{pages}{094}
  (\bibinfo{year}{2011}), \eprint{0809.2596}.

\bibitem[{\citenamefont{Son and Surowka}(2009)}]{Son:2009tf}
\bibinfo{author}{\bibfnamefont{D.~T.} \bibnamefont{Son}} \bibnamefont{and}
  \bibinfo{author}{\bibfnamefont{P.}~\bibnamefont{Surowka}},
  \bibinfo{journal}{Phys.Rev.Lett.} \textbf{\bibinfo{volume}{103}},
  \bibinfo{pages}{191601} (\bibinfo{year}{2009}), \eprint{0906.5044}.

\bibitem[{\citenamefont{Hou et~al.}(2012)\citenamefont{Hou, Liu, and
  Ren}}]{Hou:2012xg}
\bibinfo{author}{\bibfnamefont{D.-F.} \bibnamefont{Hou}},
  \bibinfo{author}{\bibfnamefont{H.}~\bibnamefont{Liu}}, \bibnamefont{and}
  \bibinfo{author}{\bibfnamefont{H.-c.} \bibnamefont{Ren}},
  \bibinfo{journal}{Phys. Rev.} \textbf{\bibinfo{volume}{D86}},
  \bibinfo{pages}{121703} (\bibinfo{year}{2012}), \eprint{1210.0969}.

\bibitem[{\citenamefont{Chen et~al.}(2013)\citenamefont{Chen, Pu, Wang, and
  Wang}}]{Chen:2012ca}
\bibinfo{author}{\bibfnamefont{J.-W.} \bibnamefont{Chen}},
  \bibinfo{author}{\bibfnamefont{S.}~\bibnamefont{Pu}},
  \bibinfo{author}{\bibfnamefont{Q.}~\bibnamefont{Wang}}, \bibnamefont{and}
  \bibinfo{author}{\bibfnamefont{X.-N.} \bibnamefont{Wang}},
  \bibinfo{journal}{Phys.Rev.Lett.} \textbf{\bibinfo{volume}{110}},
  \bibinfo{pages}{262301} (\bibinfo{year}{2013}), \eprint{1210.8312}.

\bibitem[{\citenamefont{Gao and Wang}(2015)}]{Gao:2015zka}
\bibinfo{author}{\bibfnamefont{J.-h.} \bibnamefont{Gao}} \bibnamefont{and}
  \bibinfo{author}{\bibfnamefont{Q.}~\bibnamefont{Wang}},
  \bibinfo{journal}{Phys. Lett.} \textbf{\bibinfo{volume}{B749}},
  \bibinfo{pages}{542} (\bibinfo{year}{2015}), \eprint{1504.07334}.

\bibitem[{\citenamefont{Hidaka et~al.}(2017)\citenamefont{Hidaka, Pu, and
  Yang}}]{Hidaka:2016yjf}
\bibinfo{author}{\bibfnamefont{Y.}~\bibnamefont{Hidaka}},
  \bibinfo{author}{\bibfnamefont{S.}~\bibnamefont{Pu}}, \bibnamefont{and}
  \bibinfo{author}{\bibfnamefont{D.-L.} \bibnamefont{Yang}},
  \bibinfo{journal}{Phys. Rev.} \textbf{\bibinfo{volume}{D95}},
  \bibinfo{pages}{091901} (\bibinfo{year}{2017}), \eprint{1612.04630}.

\bibitem[{\citenamefont{Gao et~al.}(2017)\citenamefont{Gao, Pu, and
  Wang}}]{Gao:2017gfq}
\bibinfo{author}{\bibfnamefont{J.-h.} \bibnamefont{Gao}},
  \bibinfo{author}{\bibfnamefont{S.}~\bibnamefont{Pu}}, \bibnamefont{and}
  \bibinfo{author}{\bibfnamefont{Q.}~\bibnamefont{Wang}},
  \bibinfo{journal}{Phys. Rev.} \textbf{\bibinfo{volume}{D96}},
  \bibinfo{pages}{016002} (\bibinfo{year}{2017}), \eprint{1704.00244}.

\bibitem[{\citenamefont{Huang et~al.}(2018)\citenamefont{Huang, Shi, Jiang,
  Liao, and Zhuang}}]{Huang:2018wdl}
\bibinfo{author}{\bibfnamefont{A.}~\bibnamefont{Huang}},
  \bibinfo{author}{\bibfnamefont{S.}~\bibnamefont{Shi}},
  \bibinfo{author}{\bibfnamefont{Y.}~\bibnamefont{Jiang}},
  \bibinfo{author}{\bibfnamefont{J.}~\bibnamefont{Liao}}, \bibnamefont{and}
  \bibinfo{author}{\bibfnamefont{P.}~\bibnamefont{Zhuang}},
  \bibinfo{journal}{Phys. Rev.} \textbf{\bibinfo{volume}{D98}},
  \bibinfo{pages}{036010} (\bibinfo{year}{2018}), \eprint{1801.03640}.

\bibitem[{\citenamefont{Yang et~al.}(2018)\citenamefont{Yang, Fang, Wang, and
  Wang}}]{Yang:2017sdk}
\bibinfo{author}{\bibfnamefont{Y.-G.} \bibnamefont{Yang}},
  \bibinfo{author}{\bibfnamefont{R.-H.} \bibnamefont{Fang}},
  \bibinfo{author}{\bibfnamefont{Q.}~\bibnamefont{Wang}}, \bibnamefont{and}
  \bibinfo{author}{\bibfnamefont{X.-N.} \bibnamefont{Wang}},
  \bibinfo{journal}{Phys. Rev.} \textbf{\bibinfo{volume}{C97}},
  \bibinfo{pages}{034917} (\bibinfo{year}{2018}), \eprint{1711.06008}.

\bibitem[{\citenamefont{Pu et~al.}(2011)\citenamefont{Pu, Gao, and
  Wang}}]{Pu:2010as}
\bibinfo{author}{\bibfnamefont{S.}~\bibnamefont{Pu}},
  \bibinfo{author}{\bibfnamefont{J.-h.} \bibnamefont{Gao}}, \bibnamefont{and}
  \bibinfo{author}{\bibfnamefont{Q.}~\bibnamefont{Wang}},
  \bibinfo{journal}{Phys. Rev.} \textbf{\bibinfo{volume}{D83}},
  \bibinfo{pages}{094017} (\bibinfo{year}{2011}), \eprint{1008.2418}.

\bibitem[{\citenamefont{Ozonder}(2010)}]{Ozonder:2010zy}
\bibinfo{author}{\bibfnamefont{S.}~\bibnamefont{Ozonder}},
  \bibinfo{journal}{Phys. Rev.} \textbf{\bibinfo{volume}{C81}},
  \bibinfo{pages}{062201} (\bibinfo{year}{2010}), \bibinfo{note}{[Erratum:
  Phys. Rev.C84,019903(2011)]}, \eprint{1004.3883}.

\bibitem[{\citenamefont{Chen et~al.}(2014)\citenamefont{Chen, Son, Stephanov,
  Yee, and Yin}}]{Chen:2014cla}
\bibinfo{author}{\bibfnamefont{J.-Y.} \bibnamefont{Chen}},
  \bibinfo{author}{\bibfnamefont{D.~T.} \bibnamefont{Son}},
  \bibinfo{author}{\bibfnamefont{M.~A.} \bibnamefont{Stephanov}},
  \bibinfo{author}{\bibfnamefont{H.-U.} \bibnamefont{Yee}}, \bibnamefont{and}
  \bibinfo{author}{\bibfnamefont{Y.}~\bibnamefont{Yin}},
  \bibinfo{journal}{Phys. Rev. Lett.} \textbf{\bibinfo{volume}{113}},
  \bibinfo{pages}{182302} (\bibinfo{year}{2014}), \eprint{1404.5963}.

\bibitem[{\citenamefont{Duval et~al.}(2015)\citenamefont{Duval, Elbistan,
  Horvathy, and Zhang}}]{Duval:2014cfa}
\bibinfo{author}{\bibfnamefont{C.}~\bibnamefont{Duval}},
  \bibinfo{author}{\bibfnamefont{M.}~\bibnamefont{Elbistan}},
  \bibinfo{author}{\bibfnamefont{P.~A.} \bibnamefont{Horvathy}},
  \bibnamefont{and} \bibinfo{author}{\bibfnamefont{P.~M.} \bibnamefont{Zhang}},
  \bibinfo{journal}{Phys. Lett.} \textbf{\bibinfo{volume}{B742}},
  \bibinfo{pages}{322} (\bibinfo{year}{2015}), \eprint{1411.6541}.

\bibitem[{\citenamefont{Stone et~al.}(2015{\natexlab{a}})\citenamefont{Stone,
  Dwivedi, and Zhou}}]{Stone:2014fja}
\bibinfo{author}{\bibfnamefont{M.}~\bibnamefont{Stone}},
  \bibinfo{author}{\bibfnamefont{V.}~\bibnamefont{Dwivedi}}, \bibnamefont{and}
  \bibinfo{author}{\bibfnamefont{T.}~\bibnamefont{Zhou}},
  \bibinfo{journal}{Phys. Rev.} \textbf{\bibinfo{volume}{D91}},
  \bibinfo{pages}{025004} (\bibinfo{year}{2015}{\natexlab{a}}),
  \eprint{1406.0354}.

\bibitem[{\citenamefont{Stone et~al.}(2015{\natexlab{b}})\citenamefont{Stone,
  Dwivedi, and Zhou}}]{Stone:2015kla}
\bibinfo{author}{\bibfnamefont{M.}~\bibnamefont{Stone}},
  \bibinfo{author}{\bibfnamefont{V.}~\bibnamefont{Dwivedi}}, \bibnamefont{and}
  \bibinfo{author}{\bibfnamefont{T.}~\bibnamefont{Zhou}},
  \bibinfo{journal}{Phys. Rev. Lett.} \textbf{\bibinfo{volume}{114}},
  \bibinfo{pages}{210402} (\bibinfo{year}{2015}{\natexlab{b}}),
  \eprint{1501.04586}.

\bibitem[{\citenamefont{Huang and Sadofyev}(2018)}]{Huang:2018aly}
\bibinfo{author}{\bibfnamefont{X.-G.} \bibnamefont{Huang}} \bibnamefont{and}
  \bibinfo{author}{\bibfnamefont{A.~V.} \bibnamefont{Sadofyev}}
  (\bibinfo{year}{2018}), \eprint{1805.08779}.

\bibitem[{\citenamefont{Chen et~al.}(2015)\citenamefont{Chen, Son, and
  Stephanov}}]{Chen:2015gta}
\bibinfo{author}{\bibfnamefont{J.-Y.} \bibnamefont{Chen}},
  \bibinfo{author}{\bibfnamefont{D.~T.} \bibnamefont{Son}}, \bibnamefont{and}
  \bibinfo{author}{\bibfnamefont{M.~A.} \bibnamefont{Stephanov}},
  \bibinfo{journal}{Phys. Rev. Lett.} \textbf{\bibinfo{volume}{115}},
  \bibinfo{pages}{021601} (\bibinfo{year}{2015}), \eprint{1502.06966}.

\bibitem[{\citenamefont{Kharzeev
  et~al.}(2016{\natexlab{b}})\citenamefont{Kharzeev, Stephanov, and
  Yee}}]{Kharzeev:2016sut}
\bibinfo{author}{\bibfnamefont{D.~E.} \bibnamefont{Kharzeev}},
  \bibinfo{author}{\bibfnamefont{M.~A.} \bibnamefont{Stephanov}},
  \bibnamefont{and} \bibinfo{author}{\bibfnamefont{H.-U.} \bibnamefont{Yee}}
  (\bibinfo{year}{2016}{\natexlab{b}}), \eprint{1612.01674}.

\end{thebibliography}

\end{document}